\documentclass[prc,preprint,nofootinbib,preprintnumbers,
superscriptaddress,tightenlines,longbibliography]{revtex4-1}

\usepackage{amsmath,amsfonts,mathrsfs}    
\usepackage{graphicx}   
\usepackage{subfigure}
\usepackage{color}
\usepackage[colorlinks=true,linkcolor=blue,citecolor=blue, urlcolor=blue]{hyperref}   
\usepackage{bm} 
\usepackage[inline]{enumitem}
\usepackage{soul}
\usepackage{graphicx}
\usepackage{amsmath,amssymb}
\usepackage{soul}
\usepackage{textcomp}
\usepackage{hyperref}
\usepackage{subfigure}
\usepackage[toc,page]{appendix}

\def\half{{\textstyle\frac{1}{2}}}

\def\x{{\bm x}}

\def\q{{\bm q}}
\def\k{{\bm k}}

\def\mzero{m_0}
\def\mfrak{{\mathfrak m}}

\def\q{{\bm q}}

\def\k{{\bm k}}
\def\F{{\mathcal H}}
\def\W{{S}}

\def\st{\begin{equation}}
\def\stp{\end{equation}}
\def\bg{\begin{eqnarray}}
\def\nd{\end{eqnarray}}
\def\Eq#1{eq.~(\ref{#1})}

\def\eq#1{(\ref{#1})}
\def\app#1{Appendix~\ref{#1}}
\def\Fig#1{Fig.~\ref{#1}}

\def\Sect#1{Sect.~\ref{#1}}
\def\Ref#1{Ref.~\cite{#1}}
\def\llangle{\left\langle}
\def\rrangle{\right\rangle}
\def \bes {\begin{subequations}}
\def \ees {\end{subequations}}

\def \one{\mathbb I}

\def \F
\def\chemconst{{\chi_0}}

\def\sign{{\Phi}}
\def\bsigma{\bar{\sigma}}

\newcommand{\ctmfrak}{\left(\tfrac{1}{2} c_s^2 \mfrak^2 \right)}

\begin{document}

\title{Soft pions and transport near the chiral critical point}

\author{Eduardo Grossi}
\email[]{eduardo.grossi@stonybrook.edu}
\affiliation{Center for Nuclear Theory, Department of Physics and Astronomy, Stony Brook University, Stony Brook, New York 11794, USA}
\author{Alexander Soloviev}
\email[]{alexander.soloviev@stonybrook.edu}
\affiliation{Center for Nuclear Theory, Department of Physics and Astronomy, Stony Brook University, Stony Brook, New York 11794, USA}
\author{Derek Teaney}
\email[]{derek.teaney@stonybrook.edu}
\affiliation{Center for Nuclear Theory, Department of Physics and Astronomy, Stony Brook University, Stony Brook, New York 11794, USA}
\author{Fanglida Yan}
\email[]{yan.fanglida@stonybrook.edu}
\affiliation{Center for Nuclear Theory, Department of Physics and Astronomy, Stony Brook University, Stony Brook, New York 11794, USA}

\date{\today}

   \begin{abstract}
      \noindent{\bf Background:} During the expansion of a heavy ion collision, the system passes close to the $O(4)$ critical point of QCD, and thus the fluctuations of the order parameter $(\sigma, \vec{\pi})$ are expected to be enhanced. 
         {\bf Purpose:} Our goal is to compute how these enhanced fluctuations modify the transport coefficients of QCD near the pseudo-critical point. 
         We also make a phenomenological estimate for how  chiral fluctuations could effect the  momentum spectrum of soft pions.
         {\bf Method:} We first formulate the appropriate  stochastic hydrodynamic equations close to the $O(4)$ critical point.  Then, working in mean field, we  determine the correlation functions of the stress tensor and the currents which result from this stochastic real time theory, and use these  correlation functions to determine the scaling behavior of the transport coefficients. The  hydrodynamic theory also describes the propagation of pion waves, fixing the scaling behavior of the dispersion curve of soft pions.
         {\bf Results:} We present scaling functions for the shear viscosity and the charge conductivities near the pseudo-critical point, and estimate the absolute magnitude of the critical fluctuations to these parameters and the bulk viscosity.  
         Using the calculated pion dispersion curve, we estimate the expected critical enhancement of soft pion yields, and this estimate provides a plausible explanation for the excess seen in experiment relative to ordinary hydrodynamic computations.
         {\bf Conclusions:}  Our results motivate further phenomenological and numerical work on the implications of chiral symmetry on real time properties of thermal QCD near the pseudo-critical point. 

   \end{abstract}


\maketitle

\clearpage

\section{Introduction}
\label{intro}

Measurements on heavy ion collisions at the Relativistic Heavy Ion Collider (RHIC) and Large Hadron Collider (LHC) are remarkably well described by viscous hydrodynamics, which predicts the measured flow harmonics and their correlations in exquisite detail~\cite{Jeon:2015dfa,Heinz:2013th}.  These hydrodynamic simulations are based on a  theory  of ordinary hydrodynamics, which ignores  chiral  symmetry breaking at low temperature and the associated chiral phase transition. 
This is reasonable at finite quark mass, where chiral symmetry is always explicitly broken. Nevertheless, if the quark mass is small enough, one would expect that 
the pattern of chiral symmetry breaking would provide a useful organizing principle for hydrodynamics, increasing its predictive power.

As a starting point for this reorganization, let us describe the appropriate hydrodynamic theory in the limit 
of two exactly massless quark flavors. 
In this limit the symmetry group of the microscopic theory is $U(1) \times SU_L(2)\times SU_R(2)$.  At high temperatures  where the symmetry of the Lagrangian is reflected in the symmetry of the thermal state,  the hydrodynamic variables are  simply the conserved charges $Q$, i.e. the energy and momentum, the iso-vector and iso-axial-vector charges, and the baryon number.
At low temperatures the symmetry of the thermal state is spontaneously broken to $U(1) \times SU_V(2)$, and the three massless Goldstone modes associated with broken symmetry (the pions) must be added  to the original list of of hydrodynamic variables, $\{Q,\pi\}$~\cite{Son:1999pa}. The theory in this case is akin to a non-abelian superfluid.
The hydrodynamic theories at high and low temperatures are separated by the chiral critical point, which is somewhat analogous to the critical point separating the normal and superfluid phases of helium~\cite{Rajagopal:1992qz,Son:2001ff}. At the critical point the hydrodynamic variables  consist of the conserved charges $Q$ and a four component  order parameter field $\Sigma \sim \langle \bar q_Rq_L \rangle $.
For $T \gg T_c$,  $\Sigma$ can be consistently integrated out, leaving an ordinary fluid state with only the conserved charges, while for $T\ll T_c$ the phase of $\Sigma$ fluctuates, reducing the hydrodynamics to a superfluid theory consisting of conserved charges and the Goldstone modes,  $\{Q,\pi\}$.

In the presence of a small but finite quark mass the theory is only approximately invariant under $SU_L(2) \times SU_R(2)$. The iso-axial vector charge is only approximately conserved, and the $\pi$ fluctuations are only
approximately massless. In addition, the system never passes directly through
the chiral phase transition, and the correlation length remains finite, but 
large. Thus at large enough distances, the theory asymptotes to ordinary hydrodynamics,
and the usual approach based on ordinary hydrodynamics is fine.
However, at shorter distances (but still macroscopic) the fluctuations of the order parameter $\Sigma$ need to be taken into account to accurately model the system with hydrodynamics.  The thermal
fluctuations of the $\Sigma$ field are incorporated into the equation of state and the transport coefficients of the
ordinary fluid theory.  By writing down the hydrodynamics theory including the
$\Sigma$, and then integrating out these modes, one can precisely determine how the critical modes affect the equation of state, and modify the  transport coefficients of the ordinary theory, such as the shear and bulk viscosities.  
This computation will determine the behavior of these parameters in the vicinity of the chiral critical point.  Our goal in this paper to perform this computation, albeit in a mean-field approximation.

The validity of the approach relies on the smallness of quark mass and the proximity of the $O(4)$ critical point in  real world QCD.  We are encouraged by Euclidean lattice QCD simulations \cite{Ding:2019prx,Kaczmarek:2020sif} at the physical pion mass  and smaller, which show that aspects of  QCD thermodynamics, such as the chiral susceptibility,  can be qualitatively, and even quantitatively, understood using $O(4)$ scaling functions. These scaling functions dictate the behavior of the singular part of the temperature dependence (at fixed quark mass) of the equation of state near the pseudo-critical point.  It seems reasonable to expect that the real time $O(4)$ scaling functions can be used to prescribe the temperature dependence of the transport parameters in the critical region with similar precision. 

The singular parts of the equation of
state can be determined by simulating an appropriate $O(4)$ symmetric
Landau-Ginzburg field theory on a 3D lattice.
In effect, this means 
that the singular part is captured by a classical effective field theory (EFT) describing the equilibrium
fluctuations of a classical order parameter field. In practice, the
classical EFT is replaced by  a spin model
and lattice techniques are used to determine the scaling functions with high precision~\cite{Engels:2014bra,Engels:2011km,Engels:2009tv}. For dynamical quantities the appropriate classical real time EFT is stochastic hydrodynamics~\cite{Hohenberg:1977ym}.  The hydrodynamic equations of motion were written down many years ago in an insightful paper by Wilczek  and Rajagopal~\cite{Rajagopal:1992qz}. We will present a somewhat different derivation of their equations of motion in \Sect{hydro}. A useful phenomenological
model which tracks the amplitude of the chiral condensate (but not the phase)  within hydrodynamics was presented in~\cite{Nahrgang:2011mv}.

A numerical simulation of the critical theory could be used to find the two point functions of the conserved currents, which in turn determine the scaling functions for the transport coefficients near the critical point. 
In the current paper we will work in a mean field approximation, in order to get a qualitative understanding for the expected scaling functions from such simulations, and to estimate the absolute magnitude of critical contributions from the $\Sigma$ field to the transport coefficients. We will reserve a numerical simulation for future work.

Currently, there is no experimental evidence
for the 
long wavelength fluctuations of the chiral condensate, which are the hallmark of the chiral phase transition\footnote{Note, however, that there is an observed enhancement of thermal dileptons in a specific mass range, which 
   can be taken
as evidence that the vector and axial-vector correlation functions are becoming degenerate, as expected when chiral symmetry is partially restored~\cite{[For a review see: ][{}]Rapp:2009yu}.}.  As a first attempt to remedy the situation,
the current paper will point out an enhancement of soft pions seen in the experimental data, and
recall that such an enhancement is an expected signature of the $O(4)$ critical point. An estimate for the magnitude of the enhancement expected from critical fluctuations encourages us to explore this explanation for the observed excess in future work.
In addition, the proposed upgrade to the ALICE detector~\cite{Colella:2019ftc} will be more sensitive to low $p_T$ pions, and this new experimental thrust provides us with additional encouragement. 

This paper builds upon our earlier work \cite{Grossi:2020ezz}, which computed the contributions of soft pions to the transport coefficients of QCD in the broken phase, and then estimated how these contributions would evolve as one approaches the critical point from below. We will recover these earlier results as a low temperature limit of the more general expressions presented here. However, while the current paper works with mean field theory, the previous results  are more general and are expected match the  full numerical simulations of stochastic hydrodynamics.

An outline of the paper is as follows: to set notation we will first
describe the thermodynamics of the $O(4)$ scaling theory, and compare
results from previous numerical simulations with the mean field
expectations used in this work. Then in \Sect{hydro}, we will provide a
general formulation of the hydrodynamic equations of motion,  and compute
the linearized propagators for the theory. These propagators will then be
used in \Sect{transport-coeff} to compute the scaling behavior of the
transport coefficients in a mean field approximation, and the results
are analyzed. Finally, in \Sect{sec:outlook} we estimate the enhanced
yield of soft pions near the chiral critical point and outline future
directions.


\section{Thermodynamic preliminaries}\label{sec:mean field}

\subsection{The magnetic equation of state at mean field}


The order parameter of the chiral phase transition is a four component field $\phi_a$ 
transforming in the defining representation of $O(4)$,
and reflects the fluctuations 
of the chiral condensate,
$\Sigma(x) \equiv - \bar q_R q_L(x) /F^2_0$ where $F_0$ is the vacuum pion decay constant. $\Sigma$ is expanded in terms of the four component field%
\footnote{ 
Roman indices  at the beginning of the alphabet $a,b,c\ldots$ 
are $O(4)$ indices. 
Isospin indices are denoted as $s,s', s'',\ldots $ etc, and are 
notated with a vector $\vec{\pi}$.
Minkowski indices are $\mu, \nu, \rho, \ldots$ etc, while 
spatial indices are $i,j,k, \ldots$. 
To lighten the notation, contraction of flavor indices are denoted by a dot, e.g. $H \cdot \phi = H_a \phi_a$ and $\mu \cdot n = \mu_{ab} \cdot n_{ab}$. More explicitly, the chiral condensate is $\left[\Sigma\right]^{\ell_1}_{\;\ell_2} =-\bar q_{R\ell_2} \, q_{L}^{\ell_1}(x)/F_0^2$, where $q^\ell=(u,d)$, and $\Sigma$ transforms as $\Sigma \rightarrow g_L \Sigma g_R^\dagger$ under a chiral rotation. 
} 
\st
\Sigma \equiv  \phi_a \tau_a = 
\sigma\, \one  + i\vec{\pi} \cdot  \vec{\lambda} \, ,
\stp
where the matrices of the Clifford algebra  $\tau_a= (\one, -i\vec{\lambda})$ 
are an amalgamation of 
the unit matrix and
the Pauli matrices,  $\vec{\lambda}$,  transforming together as a vector under $O(4)$.  
The components of $\phi_a$ are the
sigma and pion fields
\st
\label{phirelation}
(\phi_a) \equiv (\sigma, - \vec{\pi})  \, ,
\stp
where the minus sign appearing in \eqref{phirelation} is a slightly inconvenient convention.  Given the approximate $O(4)$ symmetry of the microscopic theory, there are approximately conserved charge densities, 
$n_{ab}$, transforming as an antisymmetric tensor under $O(4)$.  $n_{i j}$ is the conserved iso-vector charge, while $n_{0 i}$ is the partially conserved iso-axial-vector charge. The associated chemical
potential is $\mu_{ab}$, and we also adopt the notation $\mu^2  = \mu_{ab} \mu_{ab}$.

Close to the critical point, the Euclidean action
that determines the fluctuations in the order parameter $\phi_a$ 
at fixed temperature $T$ and chemical potential $\mu_{ab}$ is 
\begin{align} 
   \label{MFfreeenergy}
   \W_E=&  \beta \int d^3x\,  \left(p_0(T) + \frac{1}{2} \chi_0\mu^2 - \frac{1}{2} \partial_i \phi_a  \, \partial^i\phi_a -   V(\sign)+ H_a  \phi_a\right) \, ,
\end{align}
where the scalar potential is of Landau-Ginzburg form
\st
   V(\sign) = \frac{1}{2} \mzero^2(T) \sign^2 + \frac{\lambda}{4} \sign^4 \, . 
\stp
Here we have defined 
\st
\sign \equiv  \sqrt{ \phi_a \phi_a} \, , 
\stp
and 
\st
\mzero^2(T) \equiv \mfrak^2\,  \frac{(T - T_c)  }{T_c}\equiv \mfrak^2 t,
\stp
where $t$ is the reduced temperature,  and $\mfrak$ is of order the vacuum sigma mass or higher and is a constant. $H_a \equiv (H, 0, 0, 0)$ is the applied magnetic field or quark mass. At this point $T$ and $\mu$  are simply constants but have been brought inside the integral in \Eq{MFfreeenergy} to motivate 
the hydrodynamic analysis of \Sect{hydro}, where $T,\mu$ depend slowly 
space and time. 

The full partition function  takes the form
\st
Z = \int D\phi \, e^{\W_E[\phi,H]} \, , 
\stp
and reproduces the critical behavior of the equation of state. 
In spite of its well known shortcomings,
we will work in a mean field approximation. 
  The mean field takes the form
\st
\llangle \phi_a \rrangle = (\bsigma, 0 ) \, ,
\stp
where $\bsigma$ is the real solution to
\st
\label{eq:sigmabar}
\mzero^2(T) \, \bar \sigma  + \lambda\, \bsigma^3 - H = 0 \, .
\stp
It is straightforward to show that 
the solution to \eqref{eq:sigmabar} takes the scaling form
\st
\bar{\sigma} = \frac{\mfrak}{\sqrt{\lambda}} \, h^{1/3} f_G(z)\, ,   \qquad \text{with} \quad z=t h^{-2/3} \, , 
\stp
where we have defined reduced field 
\begin{align}
   h \equiv & \frac{H \sqrt{\lambda}}{\mfrak^3} .
\end{align}
$z$ is the mean field scaling variable, and  $f_G(z)$ is the (mean field) scaling function for the magnetic equation of state.  
As we will see in
the next section, the pion screening mass on the critical line, $z=0$, is given by\footnote{Here and below the subscript $c$, such as $m_c$ and $m_{\sigma c}$,  indicates that the quantity is being evaluated on the critical line $z=0$. Later we will introduce $v_c^2$  and $u_c^2$ (in eqs. \eqref{eq:v2def} and \eqref{eq:u2def}). }
\st
m_c^2  = \mfrak^2 \, h^{2/3} \, , 
\stp
and is a temperature independent constant which parametrizes $h$. 
It is convenient to express all lengths in terms of $m_c$. 
The scaling variable and equation of state take the form
\st
\bsigma = \frac{m_c}{\sqrt{\lambda}} f_G(z) \, , \qquad 
z =  \frac{m_0^2(T) }{m_c^2} = \frac{\mfrak^2}{m_c^2} \frac{(T - T_c)}{T_c}\, .  
\stp
Parametrically,  $\mfrak^2/m_c^2 = h^{-2/3}$ is a large parameter, and thus $T$ must be close to $T_c$ in order to have an order one scaling variable,  $z \sim 1$.

\begin{figure}
   \centering
   \includegraphics[width=0.48\textwidth]{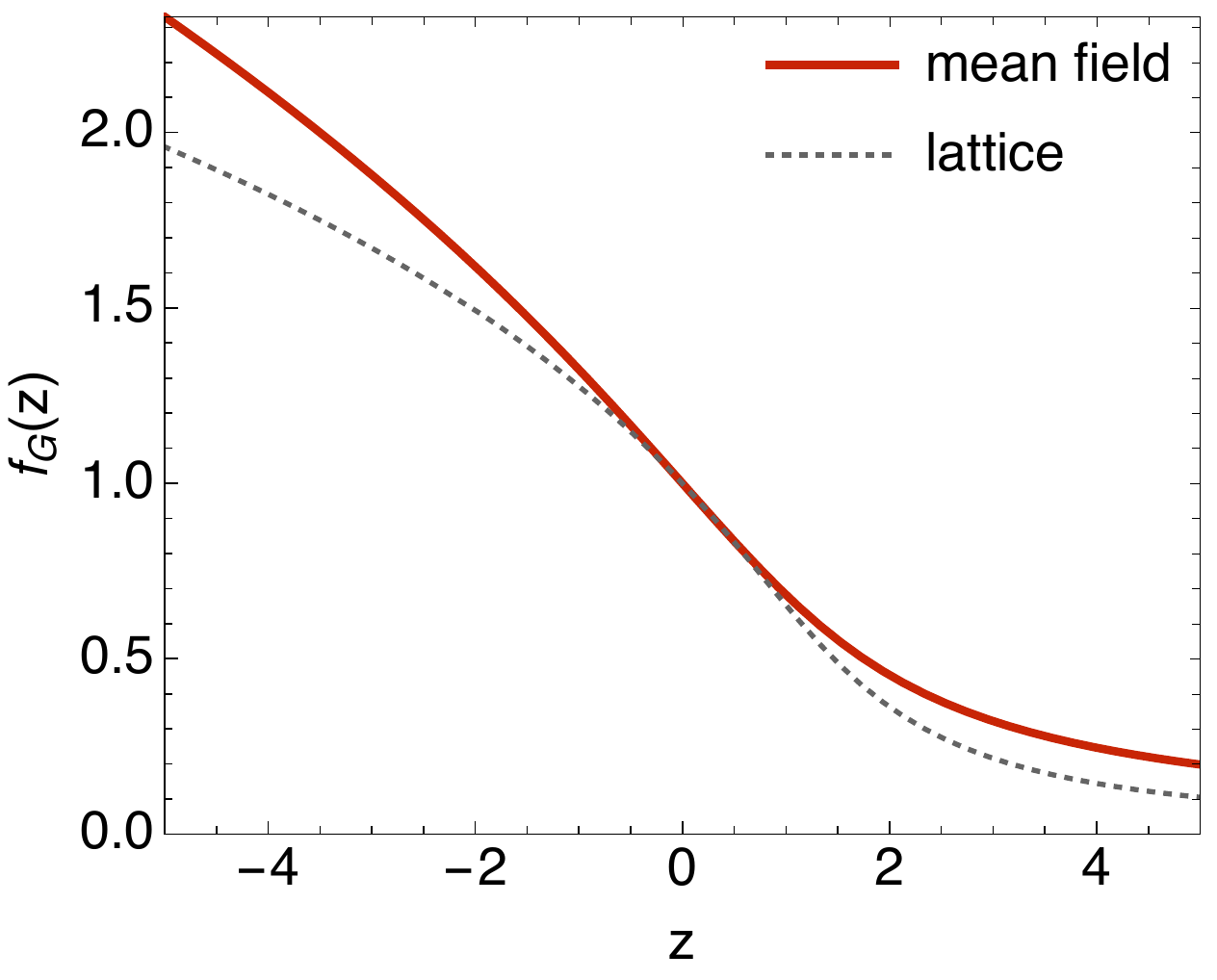}
   \hfill
   \includegraphics[width=0.47\textwidth]{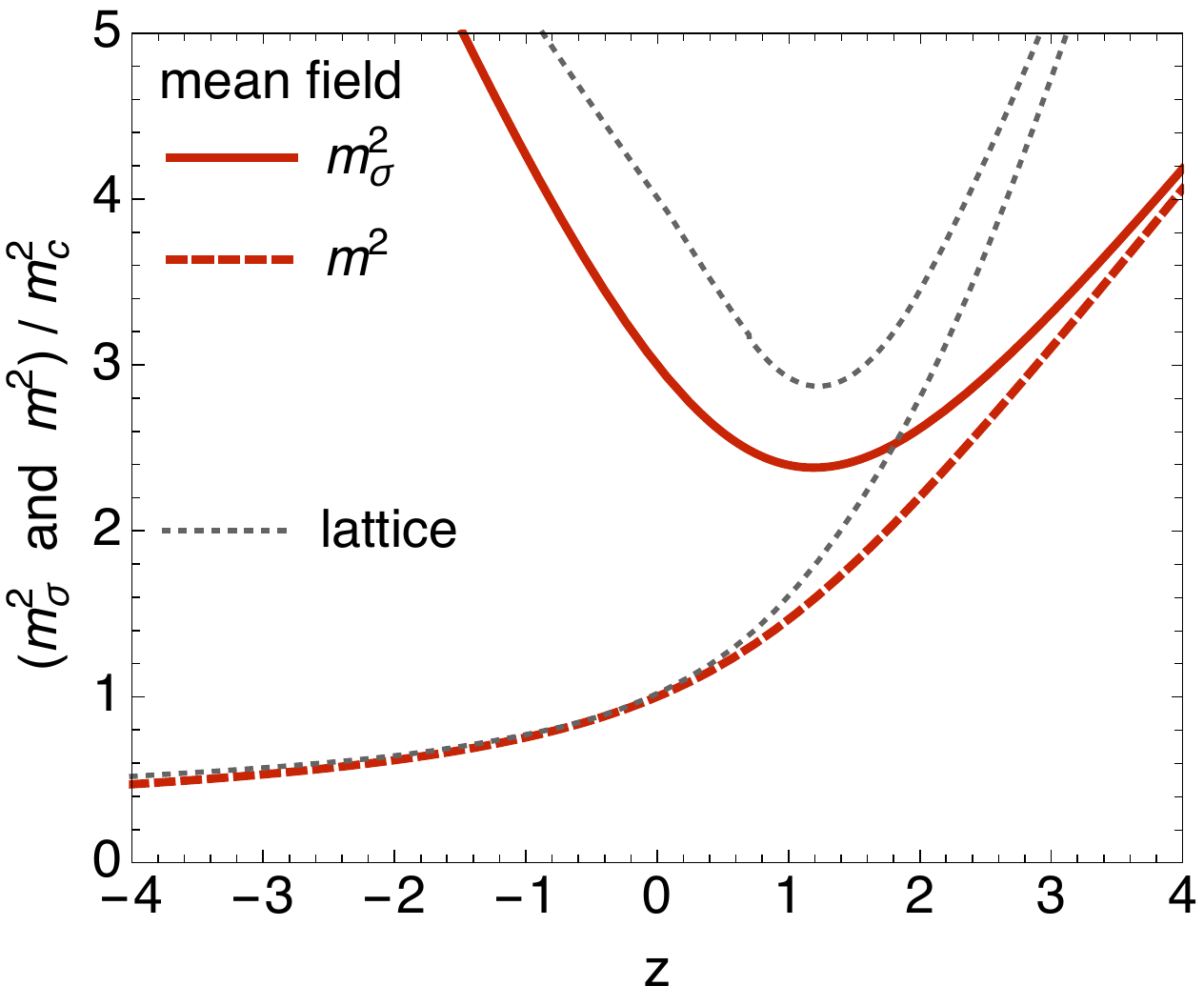}
   \caption{(a) A comparison of the mean field magnetic EOS to numerical results from lattice methods taken from~\cite{Engels:2014bra,Engels:2011km}. (b) The sigma and pion  screening masses (inverse correlation lengths), $m_\sigma$ and $m$, compared to results from lattice methods. The lattice curves  were obtained by digitizing the numerical data from Fig. 8 and Fig. 9 of \cite{Engels:2003nq}, which was subsequently fit with a parametrized form.\label{mean fieldfig}}
\end{figure}

Outside of the mean field approximation, the expectation value of the order parameter also takes 
the scaling form
\st
\bar\sigma = B\,h^{1/\delta} f_G(z)\, ,  \qquad  z = t h^{-1/\Delta} \, , 
\stp
with  $B$ a non-universal constant. $\delta$ and $\Delta$ are known 
critical exponents, and $f_G(z)$ is a known universal function~\cite{Engels:2014bra,Engels:2011km,Engels:2009tv}.
Table~\ref{tab:tableofexponents}  compares the mean field expectations
for the critical exponents to the $O(4)$ scaling theory, and \Fig{mean fieldfig}(a)  compares the mean field $f_G(z)$ to the scaling theory.
\begin{table}
   \begin{tabular}{c|c|c} 
      Exponent or ratio  & Mean field & $O(4)$ scaling theory \\ \hline
           $\beta$       &  1/2 &  0.38 \\
           $\delta$      &  3 &  4.8 \\
           $\Delta=\beta\delta$       &  3/2     & 1.83       \\
          $\nu_c=\nu/\beta\delta$ &  1/3  & 0.40 \\
          $m_{\sigma c}^2/m_c^2$ &  3  & 4.0 
 \end{tabular}
 \caption{\label{tab:tableofexponents} A comparison of mean field theory and the $O(4)$ scaling theory (see for example \cite{ParisenToldin:2003hq,Engels:2003nq} for current estimates of the $O(4)$ exponents). $m_c$ and $m_{\sigma c}$ are the pion and sigma screening masses on the critical line, $z=0$, and this ratio was taken from~\cite{Engels:2003nq}. }
\end{table}

\subsection{Static correlation functions in mean field}
\label{sec:staticorr}

Given the mean value $\bar{\sigma}$, we can 
evaluate the action \eqref{MFfreeenergy} to quadratic order 
\st
   \W_{E}  = \beta \int d^3x \, \left[p_{\sigma}(T) + \frac{1}{2} \chi_0 \mu^2   
   - \frac{1}{2} \left( \partial_i \delta \sigma \, \partial^i \delta \sigma   + m_\sigma^2 \delta \sigma^2 \right)   -  \frac{1}{2}   \left( \partial_i \vec \pi \cdot \partial^i \vec{\pi}   + m^2 \vec{\pi}^2 \right)\right],
   \stp
where  
\st
p_\sigma(T)  = p_{0}(T)  -  \left(\frac{1}{2} m_0^2(T)\bsigma^2(T) + \frac{\lambda}{4} \bsigma(T)^4 - H \bsigma(T) \right) \, .
\stp
The  sigma and pion  screening masses  are
\begin{subequations}
   \label{massdefs}
\begin{align}
         m_{\sigma}^2 &\equiv  m_c^2 \left(z + 3 f_G^2(z) \right)\, , \\
   m^2 &\equiv  \frac{H}{\bsigma(T)}   = \frac{m_c^2}{f_G(z) }\, . 
\end{align}
\end{subequations}

As in the previous section,  the screening masses (or inverse correlation 
lengths) are also defined outside of mean field theory. These 
are expected to scale as
\begin{align}
   m_{\sigma}=& {\mathfrak m }_L \,  h^{\nu_c} \, g_L(z)\, ,  \\
   m =& {\mathfrak m}_T \, h^{\nu_c} \, g_T(z)  \, , 
\end{align}
where ${\mathfrak m}_L$ and ${\mathfrak m}_T$ are  non-universal constants, and  $g_L(z)$ and $g_T(z)$  are universal scaling functions. As before,  $g_L$ and $g_T$ are normalized to unity for $z=0$. 
The ratio between  $m_\sigma$ and $m$ is also universal, 
and can be parameterized by $m_\sigma/m$ on the critical line, 
 i.e. $m_{\sigma c}^2/m_c^2$.   This universal ratio is compared to the mean field prediction of three in Table~\ref{tab:tableofexponents}. $m(z)$ and $m_{\sigma}(z)$  are extracted from the numerical work of \Ref{Engels:2003nq} and compared to mean field theory in \Fig{mean fieldfig}(b). 

The quadratic action predicts  the equal time correlation functions 
\begin{subequations}
   \label{equal-time-phi}
\begin{align}
   \label{equal-time-sigma}
   \frac{1}{V} \llangle \delta \sigma (\k)  \delta \sigma(-\k) \rrangle &= \frac{T}{k^2 + m_\sigma^2}\, ,   \\
  \frac{1}{V}  \llangle \varphi_s(\k) \varphi_{s'}(-\k) \rrangle &= \frac{T}{ \bsigma^2(k^2 + m^2)}  \delta_{ss'} \, .  
   \label{equal-time-varphi}
\end{align}
\end{subequations}
Finally,  we can use the general theory of thermodynamics fluctuations to recognize that\footnote{
   The easiest way to see this in the current framework is to recognize
   that the thermodynamic fluctuations in $n_{ab}$ are Gaussian and summed over in the  grand canonical ensemble. The factor $\tfrac{1}{4} \chi_0 \mu^2$ reflects the integration  over $n$ with the Lagrange multiplier $\mu$
   \st
   e^{\beta \int {\rm d}^3x \, \tfrac{1}{4} \chi_0 \mu^2 } = \int [Dn] \, {\rm exp}\left(- \beta \int {\rm d}^3x  \left(\tfrac{1}{4\chi_0} n^2 -  \frac{1}{2} n \cdot \mu \right) \right) \, .
   \stp
   This integral implies \Eq{gaussint}. 
}
\begin{align}
   \label{gaussint}
   \frac{1}{V} \llangle n_{ab}(\k) n_{cd}(-\k) \rrangle &=  T \chi_0  \, (\delta_{ac} \delta_{bd} - \delta_{ad}\delta_{bc}) \, .
\end{align}

Well below $T_c$,  the pion mass is small and soft pions are long lived quasi-particles~\cite{Son:1999pa,Son:2001ff,Son:2002ci}. From this context  we introduce a number of definitions following Son and Stephanov~\cite{Son:2002ci}. The phase of the condensate is 
\st
\varphi_s \equiv \frac{\pi_s}{\bsigma} \,  , 
\stp
while the associated the pion velocity squared is
\st
\label{eq:v2def}
v^2(T) \equiv   \frac{\bsigma^2(T) }{\chi_0}.
\stp
The pole mass is defined as $m_p^2 \equiv v^2 m^2$, and the soft pion 
dispersion curve takes the form 
\st
\omega^2_q = v^2 q^2 + m_p^2, 
\stp
which is parameterized by two Euclidean quantities $v^2(T)$ and $m^2(T) $.

In the next section we will develop the hydrodynamic theory for the $O(4)$ model. 
The real time correlation functions constructed from this theory will reproduce 
\eqref{equal-time-phi} and \eqref{gaussint} after integrating over frequency.

\section{Hydrodynamics }\label{hydro}

Having discussed the thermodynamics, we are ready to  derive the corresponding hydrodynamic theory.  
The resulting equations of motion are  equivalent to those 
derived previously by Rajagopal and Wilczek using Poisson bracket methods~\cite{Rajagopal:1992qz}.
Well below $T_c$,  the equations of motion resemble a non-abelian superfluid theory and also have been analyzed~\cite{Son:1999pa,Son:2002ci,Jain:2016rlz,Grossi:2020ezz}. The methodology 
here follows closely our previous work~\cite{Grossi:2020ezz}.

\subsection{Ideal hydrodynamics and the Josephson constraint }

To derive the ideal hydrodynamic expressions we follow an expedient procedure
procedure outlined in \cite{Jensen:2012jh}  and 
take as  hydrodynamic action 
\begin{equation}
   \label{eq:hydroaction}
   \W[g_{\mu\nu},A_{\mu},H]= \int d^4x  \sqrt{-g} \, p_{\Sigma}(T,\mu,(\partial_\perp\phi)^2, \phi^2, H\cdot\phi) \, , 
\end{equation}
where the redefined pressure takes the same form as its Euclidean counterpart  
\begin{equation}
   \label{redef-pressure}
   p_\Sigma(T,\mu, (\partial_\perp \phi)^2 , \phi^2, H\cdot\phi) \equiv p(T) + \frac14 \chi_0  \,\mu \cdot \mu -  \frac12 \Delta^{\mu\nu} D_{\mu}\phi \cdot  D_{\nu}\phi  -V(\sign)  + H \cdot \phi \, ,
\end{equation}
but we have replaced the integration over thermal circle with
an integration over time
\st
\beta = \int_0^{\beta} d\tau   \rightarrow \int dt \, .
\stp
We also have added external gauge and gravitational fields,  $(A_{\mu})_{ab}$ and  $g_{\mu\nu}$, for the  purpose of deriving the stress tensor and currents,   and ultimately these sources will  be set to zero.  
In these expressions $T \equiv (-\beta^{\mu}g_{\mu\nu} \beta^{\mu})^{-1/2}$, and then we define 
$u^{\mu}\equiv T\beta^{\mu}$, and $\Delta^{\mu\nu} \equiv g^{\mu\nu} + u^{\mu}u^{\nu}$. The chemical potential can be written
\st
\mu_{ab} =  \left(u^\rho \tilde \mu_\rho   + u^\rho A_{\rho}\right)_{ab} \, ,
\stp
where $(\tilde \mu_\rho)_{ab}$ is the contact chemical potential and is independent of $A_{\rho}$. 
The covariant derivative is 
\st
(D_{\mu} \phi)_a  = \partial_\mu \phi_a    - \tfrac{i}{2} (A_{\mu} \cdot \mathcal J)_{ab} \phi_b \, , 
\stp
where $\mathcal J_{cd}$ are the generators  $O(4)$ rotation group 
\st
(i \mathcal J_{cd})_{ab} = \delta_{ca} \delta_{db}-  \delta_{cb} \delta_{da} \,.
\stp

Setting the external fields to zero for simplicity,  the differential of pressure at fixed $H$  follows from the form of $p_{\Sigma}$ 
\begin{equation}
   \label{pdifferential}
   d p_\Sigma = s_{\Sigma} \, dT +\frac12 n_{ab} \, d\mu_{ab} - \frac12 d ( \partial_{\perp}^\mu \phi)^2 +\left(-\frac{\partial V}{\partial \phi_a} +H_a\right)d\phi_a \, , 
\end{equation}
which defines the entropy density, $s_\Sigma \equiv \partial p_\Sigma/\partial T$,  and the number densities, $n_{ab} \equiv   2 \, \partial p_{\Sigma}/\partial \mu_{ab}$, respectively\footnote{ 
   More explicitly,  $s_{\Sigma}(T)  = s(T) -\frac{1}{2 T_c} \mfrak^2 \, \Phi^2 \,$. The factor of two in the definition of $n_{ab}$, leading to $n_{ab} = \chi_0 \mu_{ab}$,  is a symmetry factor, i.e. $n_{ab}\equiv \partial p_\Sigma/\partial \mu_{ab} - \partial p_\Sigma/\partial\mu_{ba}$ with $\mu_{12}$ and
   $\mu_{21}$ treated as independent variables. Similar symmetry factors for symmetric and antisymmetric tensors are present in the definitions of $T^{\mu\nu}$ and $J^{\mu}_{ab}$.
   }. 
Here and below $d\equiv u^{\mu} \partial_{\mu}$ and $\partial_{\perp}^{\mu} = \Delta^{\mu\nu} \partial_\nu$.

Varying the action with respect to the metric yields the conserved stress
tensor $\partial_{\mu} T^{\mu\nu}=0$. Recognizing that both the temperature and 
the chemical potential depend implicitly on the metric 
after they are written in terms of $\beta^{\mu}$,
the variation of the action gives
\begin{align}
   T^{\mu\nu} = \left. \frac{2}{\sqrt{-g}}\frac{\delta \W}{\delta g_{\mu \nu}} \right|_{g=A=0} =(\varepsilon_\Sigma +p_\Sigma)\,u^{\mu}u^{\nu} +p_\Sigma g^{\mu\nu} +
\partial^{\mu}\phi\cdot
\partial^{\nu}\phi -u^\mu u^{\nu} (u^\sigma \partial_{\sigma} \phi)\cdot
(u^\rho \partial_{\rho}\phi),
\end{align}
where in this expression the energy density  has been defined through the Gibbs-Duhem relation 
\begin{equation}
   \label{redef-energy}
   \varepsilon_\Sigma= -p_{\Sigma} + T s_{\Sigma} + \tfrac{1}{2} \mu_{ab} \cdot n_{ab} \, .
\end{equation}

We can find the partial current conservation equation by
requiring that the action in  \eq{eq:hydroaction} be invariant under gauge transformations.  
We will limit the discussion to weak fields and switch off the gravitational field for this purpose.
Under an infinitesimal $O(4)$ rotation with parameters $\omega_{cd}(x)$,  the 
gauge fields and magnetic field transform as
\begin{subequations}
   \label{gaugerules}
\begin{align}
   A_{\mu,cd} \rightarrow& A_{\mu,cd}  +  \partial_{\mu}  \omega_{cd} \, ,   \\
   \delta H_a  \rightarrow& H_a +   \tfrac{i}{2} (\omega \cdot {\mathcal J})_{ab}   H_{b} \, . 
\end{align}
\end{subequations}
Then,  requiring invariance of the action  under the rotation, 
\begin{align}
   \delta \W  =& \int d^4x \, \frac{\delta \W}{\delta A_{\mu,ab}} \,  \delta A_{\mu,ab} + \frac{\delta \W}{\delta H_a} \delta H_a =0,
\end{align}
and inserting  the transformation rules \eqref{gaugerules}, we find
partial  current conservation 
\st
\label{eq:pcac}
\partial_{\mu} J^{\mu}_{cd} = \phi_c H_d - \phi_d H_c \, .
\stp
Here the currents are given by
\begin{align}
   \label{current-def}
J^{\mu}_{ab}  \equiv 
2\frac{\delta \W}{\delta A_{\mu,ab}}
= \chi_0 \mu_{ab} u^{\mu } + (J^{\mu}_{\perp})_{ab} \, , 
\end{align}
where the first term is the normal component and the second term is the superfluid component, given by
\st \label{current-super}
(J_{\perp})_{ab}^{\mu} = \Delta^{\mu\nu} ( \partial_{\nu} \phi_a \phi_b - \partial_{\nu} \phi_b \phi_a).
\stp

To complete the equations of motion of ideal hydrodynamics, we need to 
specify a relationship between the phase of the condensate and the chemical potential known  as the Josephson constraint.
The Josephson constraint is the requirement that the field $\phi_a$ is stationary under the evolution generated by the grand potential, $\Omega=H-\frac{1}{2}\mu_{bc}N_{bc}$, i.e. the  stability of the thermal state. This reasoning leads to  a requirement on the classical Poisson bracket between $\phi$ and $\Omega$
\begin{align}
    \{\phi_a, \Omega \}=\{\phi_a, -u^{\mu} P_{\mu} -  \half \mu_{bc} N_{bc} \}
 =0 \, .
 \end{align}
 Recalling that $P_{\mu}$ and $N_{ab}$ generate translations and rotations
 respectively  (which determines their Poisson brackets with $\phi$),  we 
 find
 \st
 \label{Josephsonconstrain}
 u^{\mu} \partial_{\mu} \phi_a
 +
\half  
(
\mu_{ab} \phi_b
- \phi_b \mu_{b a} 
) = 0 \, . 
\stp

Alternatively, but ultimately equivalently,  the Josephson constraint  can be derived by requiring entropy conservation at ideal order~\cite{Jain:2016rlz}
\st
\partial_{\mu} (s_{\Sigma} u^{\mu}) = 0.
\stp
Appendix \ref{app-entropy} uses the conservation laws together with 
the Gibbs-Duhem relation \eqref{redef-energy} and the pressure differential \eqref{pdifferential}  
to show that entropy is only conserved if the Josephson constraint is satisfied. When viscous corrections are included at subsequent orders in the gradient expansion, the Josephson constraint will need to be modified.

Finally, it is useful to express the Josephson constraint in terms of the
amplitude, $\Phi$,  and $SU(2)$ phase,  $U$.
Writing  the chiral condensate as 
\st
   \Sigma = \phi_a \tau_a = \Phi U \, , 
\stp
the Josephson constraint can be written
\begin{subequations}
\begin{align}
   u^{\mu} \partial_{\mu} \Phi &= 0 \,,  \\
   iu^{\mu} \partial_{\mu} U U^{-1} &=  
     \mu_{L} -  U \mu_R U^\dagger \, ,
\end{align}
\end{subequations}
where $\mu_{L}=\tfrac{1}{2} \mu_{ab} \tau_{ab}$ and $\mu_R = \tfrac{1}{2} \mu_{ab} \bar\tau_{ab}$ are the left and right chemical potentials\footnote{The Clifford algebra of $O(4)$ is generated by $\tau_a = (\one,-i\vec{\lambda})$ and $\bar \tau_a=(\one,i\vec{\lambda})$. The generators of the (1/2, 0) and $(0,1/2)$ representations of $O(4)$ are 
$\tau_{ab} =-i[\tau_a,\bar\tau_b]/4$ and $\bar\tau_{ab} =-i[\bar\tau_a,\tau_b]/4$, respectively.  }. The last relation between the phase and the chemical potentials is familiar from non-abelian superfluids~\cite{Son:1999pa,Grossi:2020ezz}

\subsection{Viscous corrections}

So far we have considered only the ideal equations of motion. In the dissipative case the energy-momentum tensor, the charge current, and the Josephson constraint
will acquire new terms that correspond to dissipation into the system. 
The energy-momentum tensor and the conserved currents are modified due to dissipative effects in the usual way: 
\begin{align}
\label{eq:dissipative_correction}
T^{\mu\nu}    & = T^{\mu\nu}_{\text{ideal}} + \Pi^{\mu\nu},\\
J_{ab}^{\mu} &  = J_{ab,\text{ideal}}^{\mu} + q^{\mu}_{ab}.
\end{align}
We will work in the Landau frame, where the dissipative contributions to the stress tensor and the diffusion current are taken to be orthogonal to the four velocity $u^{\mu}$,  i.e.
\begin{equation}
\Pi^{\mu\nu} u_{\mu} = 0,\quad q^{\mu}_{ab} u_{\mu} = 0.
\end{equation}
The stress tensor can be further decomposed
into a symmetric-traceless and transverse part, $\pi^{\mu\nu}$, and a trace part, $\Pi$, 
\begin{equation}
\Pi^{\mu\nu} = \pi^{\mu\nu} + \Pi \Delta^{\mu\nu} . 
\end{equation} 
In addition to the dissipative corrections to the energy-momentum tensor and the current,  the evolution equation of the chiral condensate, $\phi_a$, gets modified by dissipative effects.  Therefore it is useful to define 
\begin{equation}
u^{\mu}\partial_\mu\phi_{a} +\mu_{ab}\phi_b = \Xi_a,
\end{equation}
where $\Xi_a$ is a Lorentz scalar that  encodes the dissipative contribution to the scalar field equation of motion. 

Using the conservation of the energy-momentum tensor and the  partial conservation of the charge, the Gibbs-Duhem relation in \eqref{redef-energy}, and the pressure differential in \eqref{pdifferential} we can derive the entropy production as 
\begin{align}
\partial_\mu(s_\Sigma u^{\mu} -  \frac{\mu_{ab}}{2T}q^\mu_{ab})= 
\frac{\Xi_a}{T} \Theta_a 
- \partial_{\mu}\left(\frac{u_{\nu}}{T}\right)\Pi^{\mu\nu}
-
\partial_{\mu}\left(\frac{\mu_{ab}}{2T}\right) q^\mu_{ab}  .
\end{align}
where we have defined  the scalar quantity 
\begin{equation}
\Theta_a= \partial^2_{\perp}\phi_a -\frac{\partial V}{\partial \phi_a } +H_a ,
\end{equation}
with $\partial_\perp^2\phi_a \equiv \partial_{\mu} \partial^{\mu}_\perp \phi_a$.

Up to now we have not specified an expansion scheme; 
the equations are just a consequence of the definition of entropy, the  conservation of energy and momentum,  and the partial conservation charge. 
The positivity of  entropy production in the tensor sector can be enforced with 
\begin{equation}
\pi^{\mu\nu} = - \eta_\Sigma \sigma^{\mu\nu}, \quad\text{with} \quad \eta_\Sigma \ge0 ,
\end{equation}
where $\eta_\Sigma$ is the shear viscosity of the $O(4)$ theory.
In the vector sector we have 
\begin{equation}
q_{ab}^{\mu} = - T \sigma_\Sigma \partial^{\mu} \left(\frac{\mu_{ab}}{T}\right), \quad\text{with} \quad \sigma_\Sigma \ge0  \, , 
\end{equation} 
where $\sigma_\Sigma$ is the $O(4)$ conductivity. 

The scalar sector requires a  bit more care as there are two Lorentz scalars, $\Xi_a\Theta_a$ and $\Pi\,\partial_\mu u^{\mu}$. Generally we have as the constitutive relations for $\Pi$ and $\Xi_a$ 
\begin{align}
\Pi&=-\zeta_\Sigma \, \partial_\mu u^{\mu} - \zeta^{(1)}_\Sigma \, \phi_a \Theta_a,\\
\Xi_a&= \zeta^{(1)}_\Sigma\, \phi_a \partial_\mu u^{\mu}+ \Gamma \, \Theta_a,  
\end{align}
where $\zeta_\Sigma$ is the bulk viscosity, $\zeta^{(1)}_\Sigma$ and  $\Gamma$ are the transport coefficients regulating the dissipative effects of the scalar field dynamics. 
The positivity of the associated quadratic form is enforced if 
\begin{equation}
\zeta_\Sigma \ge0,\quad \Gamma\ge0, \quad \text{ and } \quad \zeta_\Sigma \,\Gamma - (\zeta^{(1)}_\Sigma)^2\, \phi^2 \ge 0. 
\end{equation}

Having specified  the dissipative fluxes, it is possible to write down the energy-momentum tensor, the current,  and the scalar field equation,  including the first gradient corrections. 
The scalar field obeys a relaxation-type equation where the ideal part is the Josephson constraint 
\begin{align}\label{josephson-diss}
u^{\mu}\partial_{\mu}\phi_{a} + \mu_{ab}\phi_a= \Gamma \left[
  \partial^2_{\perp}\phi_a -\frac{\partial V}{\partial \phi_a} +H_a  
\right] +\zeta^{(1)}_\Sigma \phi_a \, \partial_\mu u^{\mu}. 
\end{align}
The energy momentum tensor now includes 
dissipative contributions due to chiral condensate
\begin{align}
\Pi^{\mu\nu} = 
- \eta_\Sigma \sigma^{\mu\nu} - \Delta^{\mu\nu} \left[
\zeta_\Sigma \partial_\mu u^{\mu} - \zeta^{(1)}_\Sigma \phi_\alpha
\left(
 \partial^2_{\perp} \phi_a -\frac{\partial V}{\partial \phi_a } +H_a 
\right)
 \right] . 
\end{align}
Finally the current has the form
\begin{align}
(J^{\mu})_{ab} =
n_{ab} u^{\mu } + (J^{\mu}_\perp)_{ab} -T \sigma_\Sigma\, \Delta^{\mu\nu} \partial_{\nu}\left(\frac{\mu_{ab}}{T}\right) \, ,
\end{align}
and is partially conserved as in \eqref{eq:pcac}.
The coefficient $\zeta^{(1)}_\Sigma $ is an independent  transport coefficient that couple the expansion rate $\partial_\mu u^{\mu}$ to the Josephson constraint and vice versa.  
Near the phase transition $\phi_\alpha$ is approximately zero, which means this term is subdominant and can be neglected. 

Let us compare these equations to a number of results in 
the literature. The equations are equivalent to those of Rajagopal and Wilczek written down almost thirty years ago~\cite{Rajagopal:1992qz}; our notation for 
$\sigma_{\Sigma}$ and $\Gamma$ follows theirs.
The current reformulation is covariant and includes the coupling to the background flow. In the low temperature limit the equations match those of our previous paper~\cite{Grossi:2020ezz} (which includes a discussion of earlier work~\cite{Son:1999pa,Son:2002ci}), provided one identifies some of the coefficients\footnote{Specifically,  we have $\Gamma \rightarrow D_m$ and $\Gamma + D \rightarrow D_A$ where $D=\sigma_{\Sigma}/\chi$. }.


\subsection{Linear response} 
\label{sec:linearresponse}

Knowing the equations of motion we can determine the hydrodynamic predictions for the retarded Green functions of the system. 
In the axial channel we can consider the coupled equations of motion for $\varphi_s$ and the axial chemical potential $\mu_{0s}$,  when $\phi_a$ is
linearized  around equilibrium
\begin{align}
   \phi_a =& (\bsigma, -\bsigma  \varphi_s) \, .
\end{align}

To find the response function for $(\omega_k \varphi_s ,  \mu_{0s })$, 
we introduce  a pseudoscalar source  $H_a = (H , \delta H_s(x))$
and a gauge field $(A_0(x))_{0s}$ which are conjugate to $-\sigma \varphi_s$ and $\chi_0 \mu_{0s}$  respectively.
Due to the $O(4)$ symmetry, the external gauge field can appear in the time derivative of $\varphi_s$ and spatial gradient of the chemical potential 
\begin{subequations}
\begin{align}
\partial_t \phi_s&\to 
D_t \phi_s = \partial_t \phi_s-   (A_0)_{s0} \bsigma,\\
\partial_i\mu_{0s}&\rightarrow \partial_i \mu_{0s}- (E_i)_{0s},  
\end{align}
\end{subequations}
where $(E_i)_{0s}=(\partial_i A_0)_{0s}$. 
Applying these transformations and Fourier transforming leads us to the linearized equations in matrix form
\begin{align}
\begin{pmatrix}
-i\omega +\Gamma(k^2+m^2)& \omega_k\\
-\omega_k & -i\omega +D k^2\\
\end{pmatrix}
\begin{pmatrix}
\omega_k \varphi_s \\
\mu_{0s}
\end{pmatrix}
=
\frac{1}{\chi_0}
\begin{pmatrix}
\Gamma (k^2+m^2) & \omega_k\\
-\omega_k& D k^2\\
\end{pmatrix}
\begin{pmatrix}
-\bsigma \delta H_s/\omega_k \\
\chi_0 (A_0)_{0s}
\end{pmatrix},
\end{align}
where we have defined the diffusion coefficient $D = \sigma_{\Sigma}/\chi_0$ of the $O(4)$ symmetric theory.
The linearized equations can be solved to find the retarded propagator
\begin{multline}
\label{eq:responsefunction}
\begin{pmatrix}
\omega_k \varphi_s \\
\mu_{0s}
\end{pmatrix}
=
\frac{1}{\chi_0}\frac{1}{(-\omega^2+\omega_k^2+g_1 g_2)-i \omega \Gamma_k}\\
\times
\begin{pmatrix}
g_1(g_2-i\omega)+ \omega_k^2& -i \omega\omega_k\\
i \omega\omega_k& g_2(g_1-i\omega)+ \omega_k^2\\
\end{pmatrix}
\begin{pmatrix}
-\bsigma \delta H_s/\omega_k \\
\chi_0 \, (A_0)_{0s}
\end{pmatrix}
\, , 
\end{multline}
where, for compactness, we define the following shorthand for the dissipative rates:
\begin{subequations}
\begin{align}
   g_1 &\equiv \Gamma(k^2 +m^2)\, , \\
   g_2 &\equiv D k^2 \, , \\
   \Gamma_k &\equiv   \Gamma(k^2 + m^2) + D k^2 = g_1 + g_2 \, .
\end{align}
\end{subequations}
$\Gamma_k$ determines the damping rate of soft pions in the broken phase~\cite{Son:2002ci}.

To compute the hydrodynamic loops in the next section, it will be necessary to use the symmetrized propagator
\begin{equation} \label{symprop}
   [G_{\rm sym}(\omega)] = \frac{T }{\omega} \frac{ [G_R(\omega)] - [G_A(\omega) ]}{i}\equiv  \frac{T}{\omega} [\rho(\omega,k )] ,
 \end{equation}
 where the advanced propagator is
 \begin{align}
 [G_A(\omega)] = [G_R(\omega)]^\dagger,   
 \end{align}
 and  $\rho$ notates  the spectral density.
 Thus, the symmetrized propagator is
  \st
 \label{gsymfinal}
 [G_{\rm sym}(\omega)] =\frac{2T}{\chi_0} \frac{1}{(-\omega^2 + \omega_k^2 + g_1 g_2)^2 + (\omega \Gamma_k)^2 }
 \begin{pmatrix}
    g_1 (\omega^2 + g_2^2) +  g_2 \omega_k^2 &  -i \omega_k \omega \Gamma_k  \\
     i \omega_k \omega \Gamma_k   &   g_2 (\omega^2 + g_1^2) + g_1 \omega_k^2
 \end{pmatrix}.
 \stp

\begin{figure}
    \centering
    \includegraphics[width=0.50\textwidth]{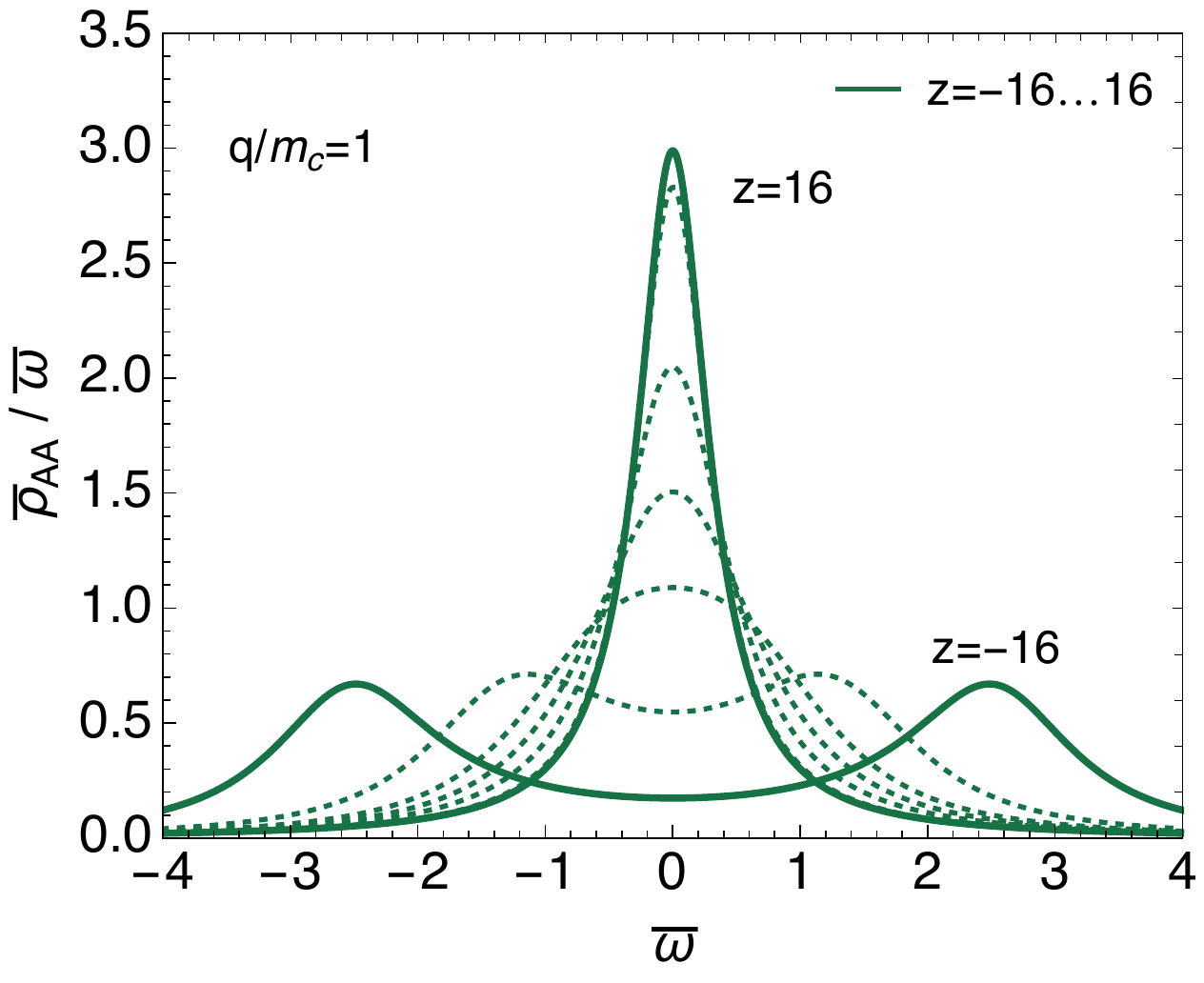} 
    \caption{\label{fig:propagator} The  spectral density $\rho_{AA}(\omega,q)$ for the axial charge density-density correlator with the scaling
 variable $z\equiv th^{-2/3}$ taking values $z=-16,-4,-1,0,1,4,16$. For
 large positive $z$ the distribution asymptotes to the simple diffusive
 pole, $\rho_{AA}/\omega \propto Dk^2/(\omega^2 + (Dk^2)^2)$, reflecting
 the diffusion of quarks. For large negative $z$ the pair of peaks reflects
 the propagating pions. We have rescaled the axes, defining $\bar{\omega}\equiv
 \omega/\Gamma m_c^2$ and $ \bar\rho_{AA}(\omega) = \rho_{AA}(\omega,q)/2
 \chi_0$, and chosen $q/m_c=1$ for illustration. For definiteness, we have
 set $D/\Gamma = 1/3$, and $v^2_c/\Gamma D m^2_c=1$, and  the motivation for these
 constants is given in the text surrounding \Eq{eq:u2def}.}
 \end{figure}

 In \Fig{fig:propagator} we exhibit the spectral density for several values of
 the scaling variable $z$ and  a specific choice of parameters discussed
 below.  It is instructive to analyze the spectral density in the pion-axial
 charge channel in two different limits, the broken phase, $z\to -\infty$, and
 the symmetric phase, $z\to  \infty$. 

In the broken phase $z\to -\infty$,  the field expectation value $\bsigma$ is large and $\omega_k \gg g_{1,2} $. 
In this limit the spectral density approaches a  Breit-Wigner form with the peaks given by the quasi-particle dispersion relation $\omega=\pm \omega_k $ and a width given by $\Gamma_k$~\cite{Son:2002ci}.
Then the denominator in the spectral density can be approximated as 
\begin{equation}
\frac{\Gamma_k}{(-\omega^2 + \omega_k^2 + g_1 g_2)^2 + (\omega \Gamma_k)^2 }\sim
\frac{1}{4  \omega_k^2 } 
\left[
   \rho(\omega,\omega_k)+ \rho(\omega,-\omega_k) 
\right] ,
\end{equation} 
where $\rho(\omega,\omega_k)$ notates the Breit-Wigner form
\begin{equation}
\rho(\omega,\omega_k) = \frac{\Gamma_k} { (-\omega+\omega_k)^2 + (\Gamma_k/2)^2} .
\end{equation}
In this limit,  we can simplify the expression of the spectral density, leading to
\begin{equation}
 \label{rhobroken}
 [\rho(\omega)] =\frac{\omega}{2 \chi_0 } \left[
\rho(\omega,\omega_k)+ \rho(\omega,-\omega_k)
\right] 
 \begin{pmatrix}
       1&  -i    \\
     i   &     1
 \end{pmatrix}+ \mathcal{O} \left( \frac{\Gamma_k}{\omega_k}\right) .
\end{equation}
Thus, there is a relation between the spectral density of pions and the axial charge\footnote{
   The response function  in \eqref{eq:responsefunction} gives the retarded function and spectral density of the chemical potential $\rho_{\mu_A\mu_A}$. Since $n_A =\chi_0 \mu_A$,  the density-density spectral function  can be obtained including the appropriate power of $\chi_0$, e.g.  $\rho_{AA}= \chi_0^2 \rho_{\mu_A\mu_A}$.
 }
\begin{align}
\rho_{AA} (\omega, k ) =  
i\chi_0 \omega_k\, \rho_{\varphi A} (\omega, k ) =
(\chi_0\omega_k)^2 \, \rho_{\varphi\varphi}  \, ,
\end{align}
which is a  manifestation of the PCAC relations. 
These relations  are the direct consequences of the Josephson equation,  $-\partial_t\varphi =n_A/\chi_0$.
Indeed, due to the ideal equation of motion for the field $\varphi_s$, we have 
\begin{align}
\rho_{AA} (\omega, k )  = - \chi_0 \, \rho_{\partial_t\varphi A} (\omega, k )  =  \chi_0^2 \, \rho_{\partial_t \varphi\partial_t\varphi}(\omega,k) \, ,  
\end{align}
which highlights that the axial charge and the time derivative of the pion field are two interchangeable concepts.

As the temperature increases, the real and imaginary parts of the propagator become of the same order of magnitude $\omega_k\sim g_1 g_2 $, 
and the parameter that governs their  relative importance  can be taken as 
\begin{equation}
   \label{eq:u2def}
u^2 = \frac{\omega^2_k}{g_1g_2 }\Big|_{k=m}  =\frac{v^2}{\Gamma D m^2}. 
\end{equation}
At the phase transition,  $z=0$, where $m= m_c$ and $v=v_c$ it is natural to assume 
that 
the real and imaginary part are the same order of magnitude~\cite{Son:2002ci}, and therefore here we will consider the case where
\begin{equation}
   \label{eq:uc2def}
u^2_c=\frac{v_c^2}{\Gamma D m_c^2}=1.
\end{equation}
The propagator also depends  on a another dimensionless parameter 
\begin{equation}
   \label{eq:r2def}
r^2=\frac{\Gamma}{\Gamma +D},
\end{equation}
which expresses the relative strengths of the axial diffusion and the order parameter relaxation.
Calculations from chiral perturbation theory found the value $r^2=3/4$~\cite{derek_tbp},  and we will adopt this number as an estimate for this order one constant.

 In the symmetric phase $z\to\infty$,  the field expectation value  is very small $ \bar \sigma \sim 0 $ and $v^2\sim 0 $. Thus, the spectral density matrix becomes diagonal
 \st
 \label{rhosymmetricphase}
 [\rho(\omega)]_{z\to \infty} =\frac{2\omega}{ \chi_0} \frac{1}{ (\omega^2 + g_2^2)(\omega^2 + g_1^2)}
 \begin{pmatrix}
    g_1 (\omega^2 + g_2^2)  &  0  \\
    0   &   g_2 (\omega^2 + g_1^2) 
 \end{pmatrix}\, ,
 \stp
 and the pion and axial charge are completely decoupled. The pion field simply relaxes to zero and the axial charge is purely diffusive. The spectral density of axial charge  is therefore 
\begin{equation}
\rho_{AA}(\omega, k ) =  \frac{
2\omega\chi_0
Dk^2}{ (\omega^2 + (D k^2)^2)},
\end{equation}
while in the pion channel  we have 
\begin{equation}
\bar \sigma^2 \rho_{  \varphi   \varphi}(\omega, k ) =  \frac{2\omega \Gamma  }{ \omega^2 + \Gamma^2  (k^2 + m^2)^2},
\end{equation}
which exhibits a simple relaxation pole with relaxation rate $\Gamma(k^2 +m^2 )$. In the symmetric phase the axial charge and the pions\footnote{In the symmetric phase there are no Goldstone bosons, the ``pions" are the pseudo-scalar fluctuations of the chiral condensate and have a very large mass. }  are completely disentangled, and their dissipative dynamics is controlled by two distinct transport coefficients. 

Moving to the $\sigma$ contribution, we see that  the linearized equation of motion is
\begin{equation}\label{sigmeq}
   \partial_t \delta \sigma = \Gamma (\nabla^2-m_{\sigma}^2 )\, \delta \sigma +\Gamma  \delta H \, , 
\end{equation}
where we have added an external source to the scalar field,  $H \rightarrow  H+\delta H$.
Solving in Fourier space, we see that the retarded Green's function is
\begin{equation}
G_{R}^{\sigma\sigma}(\omega,k)=\frac{\Gamma}{-i \omega +\Gamma (k^2  +m_{\sigma}^2) }  \, , 
\end{equation}
and the symmetrized propagator is
\begin{align}\label{sigmasym}
G_{\rm sym}^{\sigma\sigma}=\frac{2T\Gamma}{\omega^2 +\Gamma^2 (k^2  +m_{\sigma}^2)^2 }.
\end{align}
In the symmetric case ($z\to \infty$) the propagator of $\delta \sigma $ and $\varphi $ become degenerate and $O(4) $ symmetric 
\begin{equation}
\rho_{\sigma \sigma }(\omega , k ) = \bar\sigma^2 \rho_{\varphi \varphi}(\omega,k),
\end{equation}
with $m^2 = m_\sigma^2 $.

 \section{Transport coefficients}\label{transport-coeff}
 
In this section we will use the response functions calculated in
the previous section to determine the current-current and stress-stress
correlation functions. This will determine the critical 
behavior of the transport coefficients, which is analyzed and estimated
in \Sect{discussion}.

 \subsection{Hydrodynamic loops}
 \label{sec:hloops}
In the critical hydrodynamic theory we outlined in 
\Sect{hydro}, 
we have integrated out modes of order $k\sim T$.
These modes are incorporated into the transport coefficients
such as 
$\eta_{\Sigma}$ and its associated noise, $\xi_{\eta_{\Sigma}}^{\mu\nu}$.
Modes of order $k\sim m_{\sigma}$ are explicitly propagated in the theory, 
and the critical hydrodynamic theory  is defined with a cutoff  $\Lambda_{T}$
\st
         k \sim   m_{\sigma} \ll \Lambda_T \ll T.
\stp
In normal hydrodynamics, modes with $k \sim m_{\sigma}$ are 
integrated out and incorporated into the transport coefficients of the
normal theory, such as $\eta$ and its  noise. The only modes 
which are explicitly propagated are the conserved charges, and the theory 
is defined with a cutoff  $\Lambda_{\sigma}$
\st
k \ll  \Lambda_{\sigma} \ll m_{\sigma} \, .
\stp
The two transport coefficients $\eta_{\Sigma}$ and $\eta$ may be related 
by integrating out modes between $k \in [\Lambda_{\sigma}, \Lambda_T]$.

The $xy$ components of the stress tensor  in the critical hydrodynamic
theory is
\begin{align}
   T^{xy}
=& \partial^x\delta\sigma\, \partial^y\delta\sigma+\bar\sigma^2\partial^x \varphi_s\partial^y\varphi_s  + \xi^{xy}_{\eta_{\Sigma}},
\end{align}
where  the noise satisfies
\st
\label{eq:noisecorrel}
\llangle \xi_{\eta_\Sigma}^{xy}(x_1) \xi_{\eta_\Sigma}^{xy} (x_2) \rrangle 
= 2 T\eta_{\Sigma}  \delta^4 (x_1 - x_2) \, .
\stp
It is understood that the noise in the critical hydrodynamic theory is only local on scales with $k \ll \Lambda_T$, i.e. the $\delta$-function in \eqref{eq:noisecorrel} should be cutoff at the scale $\Lambda_T$ and associated with a scale-dependent parameter,  $\eta_{\Sigma}(\Lambda_T)$. 
The stress tensor in the normal hydrodynamic theory is 
simply the noise (in the absence of external flow)
\st
T^{xy}_{\rm hydro} =  \xi^{xy}_{\eta}(x)\, ,
\stp
and satisfies
\st
\llangle \xi_\eta^{xy}(x_1) \xi_{\eta}^{xy} (x_2) \rrangle 
= 2 T\eta  \delta^4 (x_1 - x_2) \, .
\stp
Matching the two effective theories yields Kubo formulas, which require that the integrated variances of the fluctuations are equal in the two theories~\cite{forster1995hydrodynamic}:
\st
2 T \eta = \int d^4x  \llangle T^{xy}_{\rm hydro} (t,\bm x) T^{xy}_{\rm hydro}(0,{\bm 0}) \rrangle  = \int d^4x \llangle T^{xy}(t, \bm x) T^{xy}(0, {\bm 0}) \rrangle \, .
\stp
Incorporating the fluctuations of $\sigma$ and $\varphi$ at one loop,  this evaluates to\footnote{
Here and below $d_A=3$ and $T_A=2$ denote the dimension and trace of the adjoint representation of the unbroken $SU(2)$ iso-vector subgroup.
The ``extra" factor of $1/\omega_k^4$ multiplying 
$G_{\rm sym}^{\varphi \varphi}$ is because of the way $G_{\rm sym}^{\varphi\varphi}$ was defined in \eqref{gsymfinal} as the symmetric correlator of $\omega_k \varphi$.
}
\begin{align}
   2 T\eta &= 2 T\eta_\Sigma(\Lambda_T)
   +2 \int^{\Lambda_T} \frac{d^3 k}{(2\pi)^3} \frac{d\omega}{2\pi}\left[
(k^x k^y G_{\rm sym}^{\sigma\sigma})^2
+d_A \frac{\bar \sigma^4}{\omega_k^4}
(k^x k^y G_{\rm sym}^{\varphi\varphi})^2\right],\label{shear-int} 
\end{align}
where we have anticipated a divergence which is regulated at the scale $\Lambda_T$,  as is appropriate for the critical hydrodynamic theory. 

The other  transport coefficients of interest here are expressed similarly as
\begin{align}
   \label{conductivity}
    2 T \sigma_{I}  =&  \int d^4x 
   \frac{1}{d_A}  \llangle \half   \{  J_{V,s}^x (t,\x), J_{V,s}^x (0, {\bm 0})  \} \rrangle,  \\
    2 T \zeta =&  \int d^4x
   \llangle \half \left\{ {\mathcal O}_{\rm bulk}(t,\x), {\mathcal O}_{\rm bulk}  
   (0, {\bm 0})  \right\} \rrangle.
\end{align}
where ${\mathcal O}_{\rm bulk } = \tfrac{1}{3} T^{i}_i + c_s^2 T^{0}_0 $. 
The bulk viscosity is significantly more complicated, and quite susceptible to physics which goes beyond the mean field approach adopted here. 
Therefore we  will evaluate the bulk viscosity only in the high temperature symmetric regime.
The relevant operators appearing in the conductivity computation and the bulk viscosity are
\begin{align}
J^x_{V,s}&=\bar\sigma^2 \epsilon_{s s^\prime  s^{\prime\prime}} \varphi_{s^\prime} \partial^x  \varphi_{s^{\prime\prime}}, \\
O_{\rm bulk,\infty} &= \frac{1}{2} c_s^2 \mfrak^2  (\delta\sigma^2 + \pi^2).  
\end{align}
Here and below we use the $\infty$ subscript to indicate
that  we have made approximations of $O_{\rm bulk}$ appropriate only 
in the symmetric  phase where $z$ is large. We have also recognized that near 
$T_c$ the terms stemming from $c_s^2 T^{0}_{0}$ are parametrically large compared
to $T^{i}_{i}$. 
Evaluating the relevant Feynman diagrams leads to 
\begin{align}
2 T \sigma_I
&=2T\sigma_\Sigma
+2 T_A \bar{\sigma}^4\int\frac{d^3 k}{(2\pi)^3} \frac{d\omega}{2\pi}\frac{1}{\omega_k^4}(k^x G_{\rm sym}^{\varphi\varphi})^2\,, \label{sigmaI-int} \\
2 T \zeta_{\infty}  &\approx 2 T\zeta_\Sigma +   2\ctmfrak^2 
\int\frac{d^3 k}{(2\pi)^3} \frac{d\omega}{2\pi}  \left[  (G_{\rm sym}^{\sigma\sigma})^2  + d_A (G_{\rm sym}^{\pi\pi})^2 \right]. \label{bulk-int}
\end{align}
The propagators in these expressions
can be read from \eqref{gsymfinal} and \eqref{sigmasym}. 
The $\sigma$ and $\pi$ propagators at large $z$ which are used in \eqref{bulk-int} are the same and are given in \eqref{sigmasym}.
To make the results of the above integrations more explicit we recall the dimensionless variables  of \Sect{sec:linearresponse}
\[
r^2=\frac{\Gamma}{\Gamma+D}, \quad \text{and} \quad
u^2=\frac{v^2}{\Gamma D m^2},
\]
and introduce the symmetric, dimensionless function $f_n(r,u)=f_n(u,r)$, defined by
\begin{align}\label{ffunc}
   f_n(r,u) &= 
   \frac{16}{15\pi}\int_0^{\infty} \frac{dk}{m}\frac{k^{2n}}{(k^2 + m^2)^3}  \frac{m^{8-2n} k^2}{  (k^2 + r^2 m^2) (k^2 + u^2 m^2) }.
   \end{align}
 For the transport coefficients in question, we will need only the following explicit expressions   
   \begin{align}
 f_3(r,u) 
   &=\frac{1}{15\left(r^2-u^2\right)}\left[\frac{r^2 \left(8 r^2+9 r+3\right)}{ (r+1)^3 }-\frac{u^2 \left(8 u^2+9 u+3\right)}{ (u+1)^3 }\right],\\
    f_2(u,r)&=\frac{1}{15 \left(r^2-u^2\right)}\left[\frac{r^2 (3 r+1)}{ (r+1)^3 }-\frac{u^2 (3 u+1)}{ (u+1)^3 }\right].
\end{align}
More details can be found in Appendix \ref{app-exact}. Then the conductivity, shear viscosity, and asymptotic bulk viscosity 
are given by
\begin{subequations}
\label{eq:transportresults}
\begin{align}
   \sigma_I(z)
&=\sigma_\Sigma +  \frac{T T_A}{32\pi m \Gamma } \left( 1-   5 u^2 (1-r^2) f_2(r,u)\right) \, ,\\
   \eta(z) &=
 \eta_\Sigma
 - 
 \frac{T }{32 \pi \Gamma} ( m_{ \sigma}  + m d_A  + m d_A 
 u^2(1-r^2) f_3 (r,u))\label{shearresult}, \\
 \zeta_{\infty}(z) &= \zeta_\Sigma   +   \frac{  T }{8\pi \Gamma m_\sigma^3 }  \ctmfrak^2 .
\end{align}
\end{subequations}
In these expressions 
the shear viscosity has been renormalized 
\begin{subequations}
\begin{align}
   \eta_{\Sigma}&=  \eta_\Sigma(\Lambda) +  \delta_{aa}\, \frac{ T \Lambda}{30 \pi^2 \Gamma},
\end{align}
\end{subequations}
and the parameters $m(z)$, $m_{\sigma}(z)$, and $u^2(z)$, depend on the 
scaling variable $z$.
\subsection{Discussion}\label{discussion}

\begin{figure}
    \centering
    \includegraphics[width=0.49\textwidth]{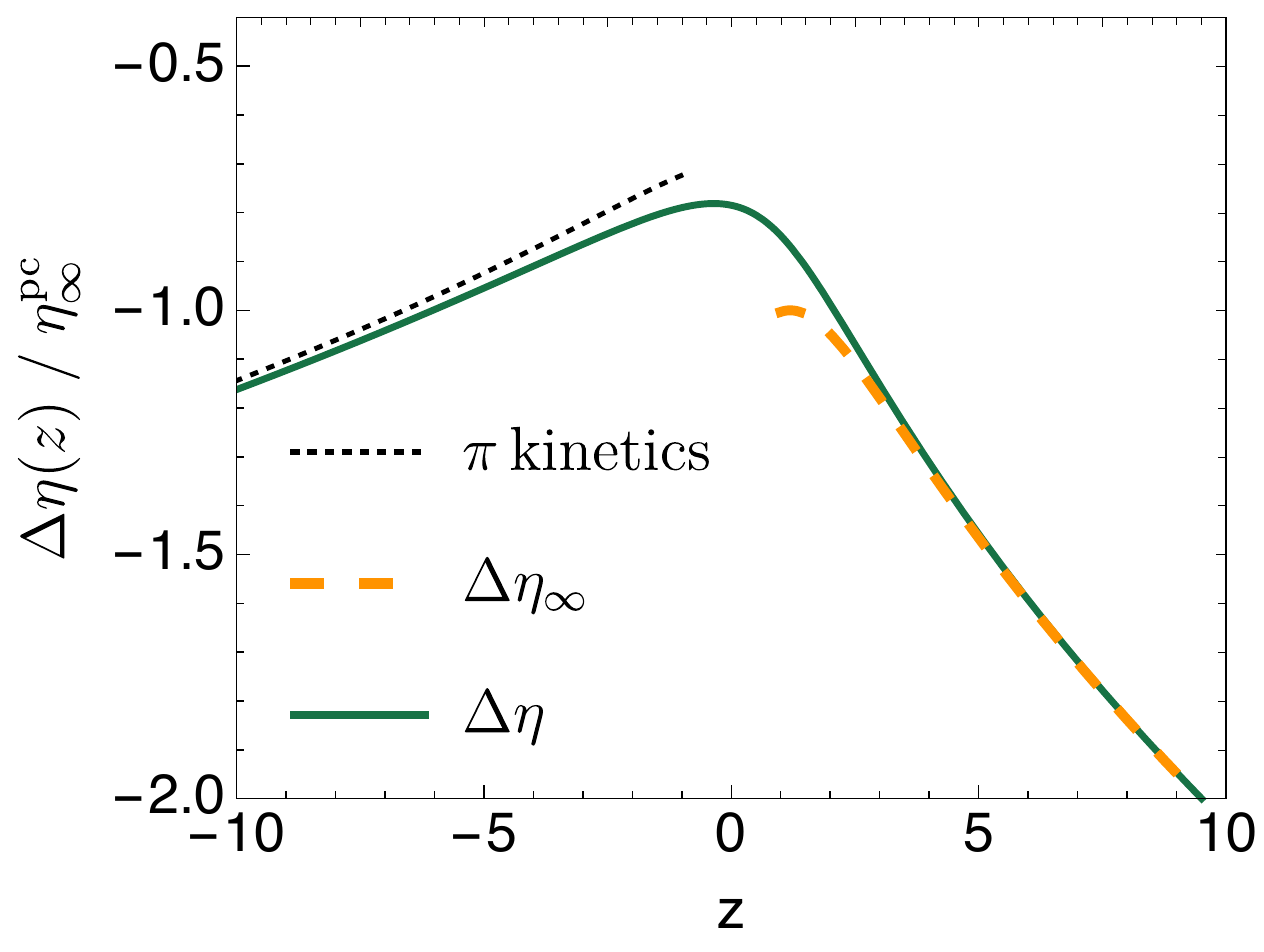} 
    \includegraphics[width=0.477\textwidth]{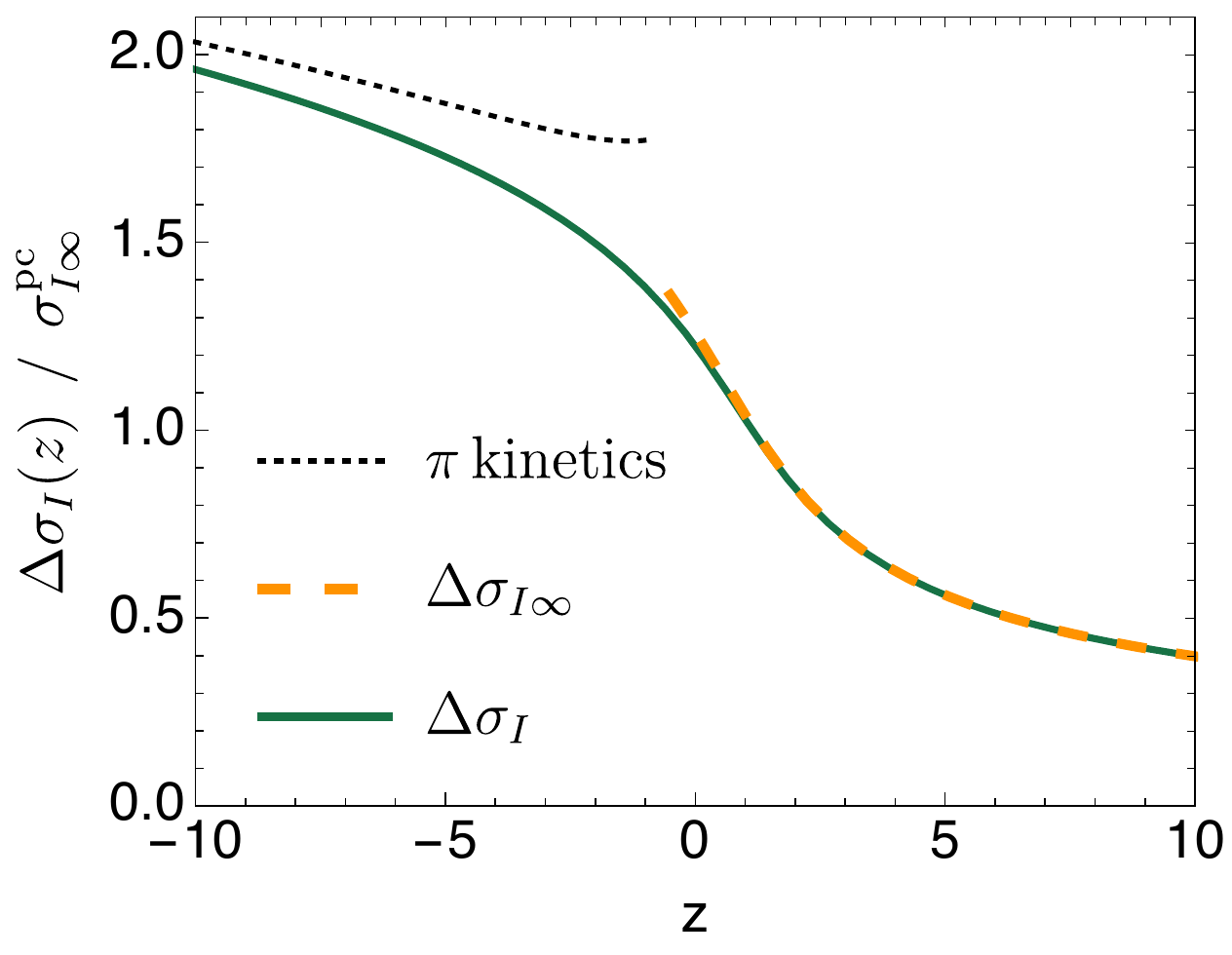} 
    \caption{The critical 
       contribution to the hydrodynamic transport coefficients, $\Delta \eta$
       and $\Delta \sigma_I$, as a function of the scaling variable $z= th^{-2/3}$. 
       The asymptotic forms at large $z$ and small $z$, e.g. $\Delta \eta_{\infty}(z)$ and ``pion kinetics" respectively, are discussed in text surrounding \Eq{eq:allinfty}.  For the  viscosity 
 we have normalized the curve by a positive constant,
 $\eta^{\rm pc}_{\infty}\equiv |\Delta \eta_{\infty}(z_{\rm pc})|$, 
 so that at the pseudo-critical point, $z_{\rm pc}=1.19$, the orange dashed asymptotic curve passes through minus one. We have defined 
 $\sigma_{I\infty}^{\rm pc}$ with an analogous notation.
 The absolute magnitudes of these normalization constants are discussed in the text surrounding \Eq{eq:finalestimates}. The curves depend weakly on two order one parameters, which we take to be $r^2$ and $u_c^2$ (see \Fig{fig:propagator}).
 \label{transport}}
 \end{figure}

 To gain an appreciation for the results of the previous section, in \Fig{transport} we plot the critical contribution to the transport coefficients $\Delta \eta$ 
 and $\Delta \sigma$ as a function of the scaling variable, $z$.  The normalization of the curves and the asymptotics  at large and small $z$ will be discussed shortly.
We emphasize that \Fig{transport} contains just the contribution from critical modes, e.g. the full shear viscosity takes the form
\st
\eta(z) = \eta_{\Sigma}  + \Delta \eta (z) \, ,
\stp
where $\eta_{\Sigma}$ is a $z$ independent constant (the regular contribution to the shear viscosity).

At large positive $z$ the propagators for the $\sigma$ and $\pi$ fields become
degenerate and take a simple form  (see \Sect{sec:linearresponse}).
This greatly simplifies the computation of the hydrodynamic loop, leading to some simple forms for the critical transport corrections. Expanding
our results in \eqref{eq:transportresults} for large $z$, or $u \rightarrow 0 $,
we find
\begin{subequations}
   \label{eq:allinfty}
\begin{align}
   \Delta \sigma_{\infty}(z) &\equiv \frac{T}{16  \pi m \Gamma } ,\\
   \Delta \eta_{\infty}(z)  &\equiv   - \frac{T m_{\sigma}}{8 \pi \Gamma}  ,
   \label{eq:etainfty} \\
   \Delta \zeta_{\infty}(z) & \equiv \frac{T  }{8\pi \Gamma m_\sigma^3} \ctmfrak^2 \label{eq:bulkinfty}.
\end{align}
\end{subequations}
These large $z$ asymptotics are presented as the (orange) dashed curves in \Fig{transport}. 
In these expressions $T$,  $\mfrak^2$, $c_s$, and $\Gamma$  are constants near $T_c$, while the remaining functions, 
$m(z)$, $m_\sigma(z)$, are scaling functions which are determined by the equilibrium magnetic equation of state.  Outside of the mean field approximation used here, $\Gamma$ is not a constant, but is expected to grow (fairly weakly) near the critical point as $\Gamma \sim m_{\sigma}^{d/2 - 2} \sim m_{\sigma}^{-1/2}$~\cite{Rajagopal:1992qz,Son:2002ci}. Treating $\Gamma$ and $D$ as constants is known in the literature as the van Hove approximation~\cite{Hohenberg:1977ym}.

\begin{figure}
 \centering
 \includegraphics[width=0.55\textwidth]{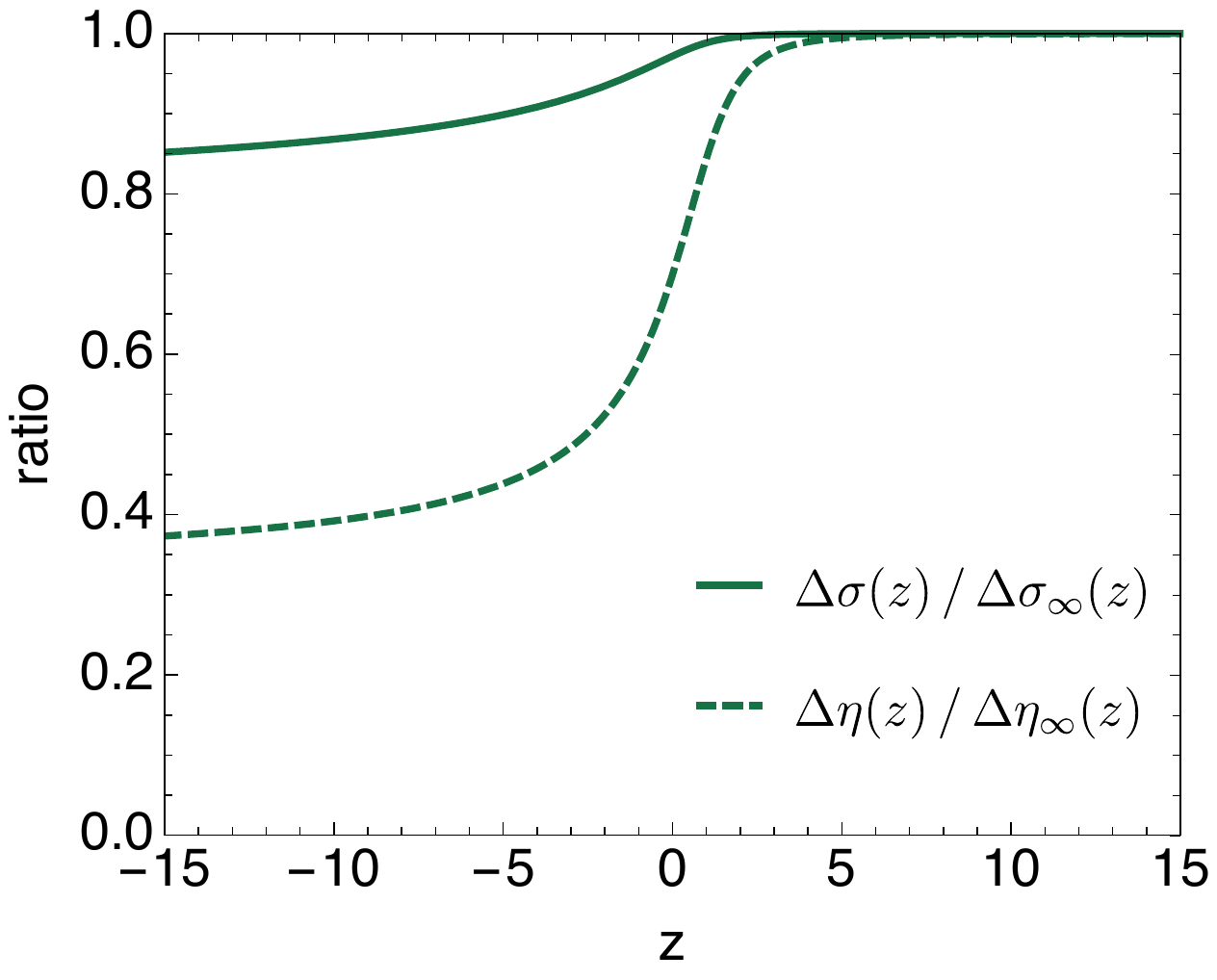} 
 \caption{Ratio of the singular part of the transport coefficients to
 the asymptotic formulas \eqref{eq:allinfty} over the full range in $z$.\label{ratioplot} }
\end{figure}

The asymptotic form of the transport coefficients sets the overall scale for our results. Thus in \Fig{transport}  we have divided each transport 
coefficient by a $z$-independent constant, the magnitude of the asymptotic result at the pseudo-critical point
\begin{subequations}
\label{eq:transportinfty}
\begin{align}
\sigma_{\infty}^{\rm pc} \equiv& \Delta \sigma_{\infty}(z_{\rm pc})  \, , \\
\eta_{\infty}^{\rm pc} \equiv& |\Delta \eta_{\infty}(z_{\rm pc}) | \, , \\
\zeta_{\infty}^{\rm pc} \equiv& \Delta \zeta_{\infty}(z_{\rm pc}) | \, .
\end{align}
\end{subequations}
Estimates for these scale coefficients in absolute units are given below. 
We also find that the simple asymptotic forms in (\ref{eq:allinfty}) provide a useful order of magnitude estimate over the whole range in $z$, and in  \Fig{ratioplot} we present the ratio between the full result and these forms. 
We expect that our asymptotic expression for the critical bulk viscosity in \eqref{eq:bulkinfty} can provide a similarly good estimate over the whole range in $z$.

For large negative $z$, the $\sigma$ is significantly heavier than the pions,  $m_{\sigma} \gg m$.
We can integrate out the heavy sigma modes and  pions with $p\sim m_\sigma$, leaving a local effective theory for soft pions with $p \sim m$. A hydrodynamic theory can be worked out for these soft pion modes coupled to the background stress~\cite{Grossi:2020ezz}. The (stochastic) hydrodynamic equations for the soft pions are equivalent to a Boltzmann equation in a ``fluid metric",  which describes how the  soft pions propagate in the background fluid~\cite{Grossi:2020ezz}. The collision terms of the kinetic equation are determined by the transport coefficients of the hydrodynamic theory.
The computation of $\sigma$, $\eta$ and $\zeta$  at large negative $z$ could thus be done in two steps: 
first one would match the hydrodynamic equations at the critical point  given in \Sect{hydro}  to the soft-pion hydro-kinetic theory, and then one would use  the soft-pion kinetic theory  to determine the transport coefficients 
as a function of the temperature. 
For large negative $z$, the results predicted by the (matched) pion  hydro-kinetic theory are shown by the black dotted curves in \Fig{transport}.  The pion kinetic theory gives a reasonable description of the results of the full theory up to its boundary of applicability, $z\sim 0$. We will use the pion kinetic theory to estimate soft pion yields in \Sect{sec:outlook}. Further details about the pion kinetic theory are given in \app{app:pionkinetics}.

Now we will make several estimates for the absolute scales
of the critical contribution to the transport coefficients, i.e. we wish
to estimate $\eta_{\infty}^{\rm pc}$, $\zeta_{\infty}^{\rm pc}$, and
$\sigma_{I\infty}^{\rm pc}$ 
defined in \eqref{eq:transportinfty}. 
These formulas have a number of physical quantities that need to be estimated, which we will do in the next paragraphs.

First, we consider the thermodynamic quantities,  which are precisely determined by lattice measurements.
To present each transport coefficient,  
we will first divide  by the corresponding susceptibilities: $sT$ in the shear and bulk cases (the momentum susceptibility),  and $T \chi_Q$  for the conductivity (the charge susceptibility). 
The pseudo-critical point is at  $T_{\rm pc}\simeq 155\,{\rm MeV}$~\cite{Borsanyi:2020fev}.
From lattice measurements of QCD thermodynamics at $T=155$ we have~\cite{Borsanyi:2010cj,Borsanyi:2013bia,Bazavov:2012jq,Bazavov:2014pvz}:
\st
sT^{-3} = 5.4\, ,  \qquad 
\chi_QT^{-2}=0.4 \, .
\stp

We will also need to estimate the screening masses, $m_\sigma$ and $m$, at
these temperatures.  
At a temperature of $T_{\rm pc}=155$,  we take from Table X of  \Ref{PhysRevD.100.094510}
\st
m_{\sigma}(T_{\rm pc}) = 0.271\,{\rm GeV}, \,  \quad \text{and} \quad m(T_{\rm pc}) = 0.198\, {\rm GeV}\, .
\stp
The mean field predictions for $m_\sigma$ and $m$ are described in
\Sect{sec:mean field};  the one free mass parameter is adjusted so that the pion
screening mass at the mean field pseudo-critical point at $z_{\rm pc}=1.19$ matches the lattice.  The corresponding mean field $\sigma$ mass at  $z_{\rm pc}=1.19$ is $m_{\sigma}=0.24\, {\rm GeV}$, which is  slightly lower than the
lattice results. 
To summarize,  in our estimates below we take
\begin{align}
m_\sigma/T = 1.56, \quad \text{and} \quad m/T=1.28 \, .
\end{align}
Finally, in order to evaluate the bulk viscosity we need to estimate $\mfrak^2$.  
In mean field theory we have\footnote{All of these relations follow
with minor algebra from \eqref{eq:sigmabar} and \eqref{massdefs}. } 
\st
\frac{\mfrak^2}{T^2} = \frac{m_{\sigma}^2}{T^2} \left(-\frac{d\log \bar \sigma}{d\log T}\right) =   \frac{m_{\sigma}^2}{T^2} \left(\frac{T_c}{T -T_c} \right) \left(- \frac{d\log f_G}{d \log z}  \right) \simeq 7.0. 
\stp
In making this estimate we have  taken
$T\simeq 155\,{\rm MeV}$ and  $T_c\simeq 132\,{\rm MeV}$~\cite{Ding:2019prx,Kaczmarek:2020sif},  and  used the mean field equation of state. In absolute units $\mfrak \simeq 0.410\,{\rm GeV}$, which seems somewhat too low  for  a cutoff scale. Indeed 
$O(4)$ fits to lattice data suggest a somewhat higher value~\cite{Kaczmarek:2020sif}. 

The real time quantities in the transport coefficients are comparatively poorly determined. The two real time parameters are the order parameter relaxation coefficient $\Gamma$ and the diffusion coefficient $D$, which set the critical 
relaxation rates,   $\Gamma m_\sigma^2\sim Dm_\sigma^2$. 
$D$ is regular near $T_c$ and determines the charge diffusion coefficient  well above $T_c$. We will therefore adopt the strong coupling estimate,  $D=1/2\pi T$~\cite{Son:2007vk,Schafer:2009dj,Heinz:2013th}, and we take $r^2 =\Gamma/(\Gamma+D)=3/4$ and $v_c^2/\Gamma D m_c^2=1$ as in \Fig{fig:propagator} (see \Sect{sec:linearresponse} for further discussion).

 
  \begin{figure}
    \centering
    \includegraphics[width=0.55\textwidth]{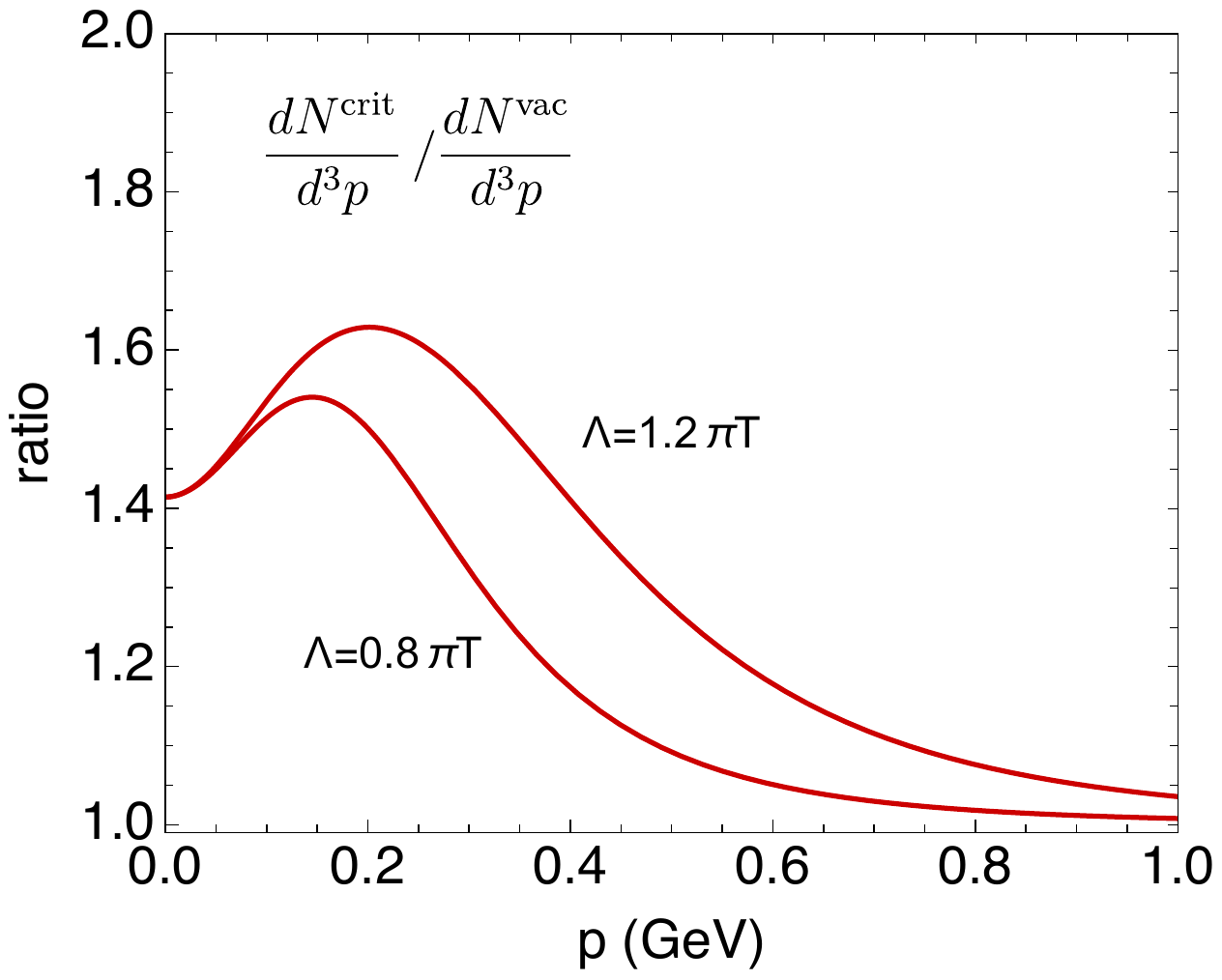} 
    \caption{The yields for soft pions due to a critical modification of the dispersion curve, relative to an expectation based on the vacuum dispersion curve (see \Eq{eq:ratioformula}). The results are shown for two different values of the cutoff $\Lambda$.  }
    \label{yields}
 \end{figure}

With these preliminaries we can estimate the scale factors for each transport coefficient. 
\begin{subequations}
   \label{eq:finalestimates}
\begin{align}
   \frac{\sigma_{I\infty}^{\rm pc}}{\chi_Q }   &= \frac{0.50}{2\pi T} \,  \left[  \left( \frac{1.5}{\pi T \Gamma} \right)  \left( \frac{1.27}{m/T} \right) \left(\frac{0.4}{\chi T^{-2}} \right) \right],\\
\frac{\eta_{\infty}^{\rm pc}}{sT}
   &= \frac{0.3}{4\pi T}  \,  \left[ \left( \frac{1.5}{\pi T\Gamma} \right) 
\left(\frac{5.4}{s/T^3} \right) \left(\frac{m_\sigma/T}{1.56}\right) \right], \\
\frac{\zeta_{\infty}^{\rm pc}}{sT}
 &=  \frac{0.025}{4\pi T} \, \left[\left( \frac{1.5}{\pi T \Gamma} \right) 
    \left(\frac{c_s^2}{0.2} \right)^2 
    \left(\frac{\mfrak^2}{7.0\, T^2} \right)^2
    \left(\frac{5.4}{s/T^3} \right) \left(\frac{1.56}{ m_\sigma/T}\right)^3
 \right] .
\end{align}
\end{subequations}
We have rescaled each transport coefficient by a value which is typical of strongly 
coupled plasmas~\cite{Son:2007vk,Schafer:2009dj}.
Thus, we see that the correction to the shear viscosity is small even in units of $1/4\pi$.
The correction to the charge diffusion coefficient $D_Q = \sigma_I/\chi_Q$ is also modest,  which is surprising given that this parameter diverges in the chiral limit.  Evidently this parametrically large enhancement does not compensate for the overall kinematics of the loop integral. 
Similarly, the bulk viscosity is also parametrically enhanced by $m_{\sigma}^{-3}$, 
but in practice this also does not compensate for other kinematic factors.
A similar observation about the bulk viscosity has been made previously in a somewhat different context~\cite{Martinez:2019bsn}.

\section{Outlook: chiral critical dynamics in heavy ion data?}
\label{sec:outlook}

The previous section estimated the influence of critical chiral modes on the transport coefficients of QCD. Given the rather large pion mass, these corrections are modest, and probably can not be observed in practice. However, it may be possible to observe the critical chiral fluctuations by directly measuring soft pions, rather than indirectly through their influence on the kinetics of the system. The approach in this section bears some similarities with~\cite{Bluhm:2020rha}, which investigated 
how a reduced chiral condensate could influence the thermal fits over a wide range of collision energies.

Current  hydrodynamic codes underestimate the yield of pions  at small transverse momenta,  see for example Fig. 3 of \cite{Devetak:2019lsk} where the data to model ratios are approximately $1.5$ for $p_T \lesssim \pi T$.
Comparable discrepancies are also found in \cite{Mazeliauskas:2019ifr,Acharya:2019yoi,Guillen:2020nul}. 
 In the broken phase, the critical dynamics is characterized by the
 formation of light Goldstone bosons,
 which is reflected in the spectral density of 
 axial charge by the formation of two quasiparticle peaks (see \Sect{sec:linearresponse}).
The
dynamics of the heavy scalar field can be neglected well below $T_c$.
With this in mind, it is reasonable to search for effects of the chiral crossover  in soft pions. Well below $T_c$,  we have previously shown that the phase-space density  of  pions with momentum $q\ll \pi T$ is approximately governed by a simple kinetic equation~\cite{Grossi:2020ezz}.  We will use this kinetic equation right at its boundary of applicability (the pseudo-critical point)
to make an estimate for the critical soft pion yield.  

The kinetic equation in the rest frame of the fluid is\footnote{For simplicity we will limit the discussion to the rest frame of the fluid, leaving the more general case to the references~\cite{Grossi:2020ezz}. }
\begin{align}
\label{eq:pionkinetic}
\frac{\partial f _\pi}{\partial t} + \frac{\partial \omega_0(q)}{\partial q_i} \frac{\partial f _\pi}{\partial x^i} - 
\frac{\partial \omega_0(q)}{\partial x^i} \frac{\partial f _\pi}{\partial q_i} 
= -\Gamma_q \left( f_\pi - \frac{T}{\omega_0(q)} \right) \,, 
\end{align} 
where $f_\pi(t,\x,\q)$ is the phase-space density of pions,  and the soft pion dispersion curve is~\cite{Son:2002ci}
\st 
\label{eq:omega0}
\omega^2_0(q)  = v^2_0 \, q^2 + m^2_{0p} \, .
\stp 
Here and in the remainder of this section we have attached the ``zero" subscript  to $v^2_0(T)$ and 
$m^2_{0p}(T)$ as a reminder that \eqref{eq:omega0} holds only for nearly zero  momenta, $q \ll \pi T$. 
The equilibrium phase-space distribution in this limit is the classical part of the Bose-Einstein distribution 
\st
\label{eq:distribution}
 f_\pi \big{\vert}_{\rm eq}  = \frac{T}{\omega_0(q)} \, .
\stp
Since $v^2_0(T)$ and $m_{0p}^2(T)$ both drop near $T_c$,  it is
natural to expect an enhancement of soft pions~\cite{Son:2001ff}. 
Here we will given estimate of this  enhancement by estimating
the critical modifications of \eqref{eq:omega0}. 

The dispersion curve in \eqref{eq:omega0} is  valid only 
for soft pions $q\ll \pi T $,  and at higher momenta 
one expects  higher derivative corrections, i.e. 
\st
\omega^2_0(q) = v^2_0 q^2 + m_{0p}^2 + \mathcal O\left(\frac{q^4}{\Lambda^2}, \frac{m_{0p}^2 q^2}{\Lambda^2} \right) \, , 
\stp
with $\Lambda \sim \pi T$.  
At large momentum the dispersion curve
should approach its vacuum form\footnote{In this formula and 
in \eqref{eq:velocity-ansatz}, $c=1$ is the speed of light and $m^2_{\rm vac} \simeq 140\,{\rm MeV}$ is the vacuum pion mass.  }
\st
\omega^2_{\rm vac}(q) = c^2q^2 + m_{\rm vac}^2  \, .
\stp
In the future,  it might be possible to constrain  the dispersion curve at fourth order in momenta using second order chiral hydrodynamics and lattice QCD measurements~\cite{Grossi:2020ezz,Son:2002ci}.
For now, we will adopt an ansatz for the pion dispersion curve at all momenta which interpolates between the  low and high momentum limits, by writing  
\st 
\label{eq:omega-interpolate}
\omega^2(q)  = v^2(q) q^2 + m_p^2(q) \, , 
\stp
with $v^2(q)$ and $m_p^2(q)$  taking the rough form
\begin{align}
   \label{eq:velocity-ansatz}
v^2(p)  =&   c^2\, (1 -F(p/\Lambda)) + v^2_0 \, F(p/\Lambda)  \, ,  \\ 
   m^2_p(p) =&  m^2_{\rm vac} (1  - F(p/\Lambda) ) + m^2_{0p}\,  F(p/\Lambda)  \, .   \label{eq:mass-ansatz}
\end{align}
Here $F(p/\Lambda)$ is any  cutoff function which has 
a Taylor series, $F(y) \simeq 1 - y^2/2 $,  at small $y$ and approaches zero for $y \sim 1$. In \Fig{yields}, we take 
\st
F(y)   = \frac{1}{1 + y^2/2 + y^4 }  \, , 
\stp
although qualitatively similar results were found with a simple cutoff,  $F(y) = {\rm max}(1 -y^2/2,0)$. 

In order to have a prediction for the dispersion curve,  
we still need to specify $v_0^2$ and $m^2_{0p}=v_0^2 m_0^2$.
These choices should be approximately consistent with lattice data on screening masses.   The lattice finds
that the pion screening mass is approximately its vacuum value for a temperature
of $135\,{\rm MeV}$, and approximately $198\,{\rm MeV}$ at the pseudo-critical point~\cite{PhysRevD.100.094510}. The temperature of $135\,{\rm MeV}$ is when the chiral susceptibility has reached approximately 60\% of its maximum and defines $z_{60}$.  In mean-field theory $z_{60} =-0.79$
and the pseudo critical point  is at $z_{\rm pc}=1.19$,  which is determined from the maximum of the  susceptibility. At $z_{60}=-0.79$, we will choose the pion's pole and screening masses to be equal to the vacuum pion mass,  and the velocity  to be $c$. The mean-field the scaling curves then dictate the screening mass at $z_{\rm pc}=1.19$, yielding:
\st
m_0(z_{\rm pc}) \simeq 0.197\, {\rm GeV} \, , 
\stp
which is nicely consistent with lattice measurements on the pion screening mass at $T=155\,{\rm MeV}$.  The same mean field scaling curves then  give the values of the pole mass and the pion velocity at the pseudo-critical point:
\begin{subequations}
   \label{eq:m0p}
\begin{align}
   m_{0p}(z_{\rm pc}) &\simeq  0.1\,{\rm GeV}\, ,  \\
   v_{0}^2(z_{\rm pc}) &\simeq 0.25 \, .
\end{align}
\end{subequations} 
In the future it would be nice to measure $v_0$ and $m$ very precisely on the lattice (they are Euclidean quantities) and to verify their critical scaling behavior in the chiral limit. 

We have now fully specified the dispersion curve $\omega^2(q)$ with eqs.~\eqref{eq:omega-interpolate}, \eqref{eq:velocity-ansatz}, \eqref{eq:mass-ansatz} and \eqref{eq:m0p}. 
Given the dispersion curve we can estimate the expected enhancement of 
yields 
\st
\label{eq:ratioformula}
\frac{ \frac{dN^{\rm crit}}{d^3p} } { \frac{dN_{\rm vac}}{d^3p } } = \frac{ \omega_{\rm vac}(p) }{\omega(p) } \, .
\stp
This prediction is shown in \Fig{yields} for two different choices of $\Lambda$.  We note that using the full Bose-Einstein distribution instead of its classical limit  $T/\omega$ produces only minor differences, which a slightly increases ratio shown in \Fig{yields}.

The ratio estimated in \Fig{yields} is roughly inline
with the observed  enhancement, although strong conclusions about the chiral critical point can not be made at this time.
Nevertheless, we find the result encouraging and it strongly motivates further research. The most obvious deficiency in our estimate is the lack of resonance decays at a naive level.  Resonances are a way of encoding interactions, and these interactions are already incorporated into the dispersion curve. It is therefore difficult ``include" resonances without double counting.   From a phenomenological perspective, it would be good to know if the fluctuations in the soft pion yield are 
correlated with rest of the pion $p_T$ spectrum, or if the variance of the soft yield has an independent component. This correlation measurement  
certainly can be done, and is ideally suited to the proposed ITS3 detector by the ALICE collaboration~\cite{ALICE}.  Additional clarifying measurements could include a direct measurement of the correlations between two soft pions.  It should be possible to provide good theoretical predictions for these correlations using $O(4)$ scaling ideas.  These predictions can be contrasted with the (presumably) rather different predictions of the hadron resonance gas. 
Finally,  it would be interesting to see if the velocity of the soft pions 
could be measured directly with non-identical particle correlations.
We hope to address these  and other topics  in the future.




\label{summary}

\begin{acknowledgments}
   We thank Anirban Lahiri and Rob Pisarski for discussions. 
This work is supported by 
the U.S. Department of Energy, Office of Science, Office of Nuclear Physics,
grants Nos.
DE\nobreakdash-FG\nobreakdash-02\nobreakdash-08ER41450. AS is supported by the Austrian Science Fund (FWF), project no. J4406.
\end{acknowledgments}

\begin{appendix}
\section{Entropy production}\label{app-entropy}
In this appendix we compute  entropy production with guidance from \cite{Bhattacharya:2011eea} and the insightful eightfold way classification scheme~\cite{Haehl:2014zda}. 
Repeating \Eq{redef-energy} and \Eq{pdifferential} for convenience, 
the entropy is given by the Gibbs-Duhem relation
\begin{equation}
   \label{eq:sdefappendix}
s_\Sigma = \frac1T (e_\Sigma +p_\Sigma  -\half \mu_{ab}n_{ab}),
\end{equation}
and the pressure differential follows from the action 
\begin{equation}
   \label{eq:dpappendix}
d p_\Sigma = s_\Sigma dT +\frac12 n_{ab} d\mu_{ab} - \frac12 d  (\partial_{\perp} \phi)^2 +\left(-\frac{\partial V}{\partial \phi_a} +H_a\right)\, d\phi_a \, . 
\end{equation}
Here $d \equiv u^{\mu} \partial_{\mu}$, and below we define $\partial u\equiv \partial_{\mu} u^{\mu}$.

Differentiating \eqref{eq:sdefappendix} and using \eqref{eq:dpappendix},  the differential of the entropy density  $ds_\Sigma $ can be written as
\begin{align}
Td s_\Sigma  
 &= de_\Sigma -\frac12 \mu_{a b}dn_{ab} -\frac12  d (\partial_{\perp}\phi)^2 +\left(-\frac{\partial V}{\partial \phi_a} +H_a\right)d\phi_a \, . 
\end{align}
The divergence of the entropy current is then:
\begin{align}
   \partial_\mu(s_\Sigma u^{\mu})&= ds_\Sigma + s_\Sigma \, \partial u \\
                                 &= \frac1T[de_\Sigma + (e_\Sigma +p_\Sigma) \partial u]
-\frac{\mu_{ab}}{2T} [dn_{ab}+ n_{ab}\partial u] 
- \frac{1}{2T} d (\partial_{\perp}\phi)^2 +\left(-\frac{\partial V}{\partial \phi_a} +H_a\right)\, \frac{d\phi_a}{T}.
\end{align}
We will now evaluate the first two terms in square brackets using
energy-momentum  and charge conservation respectively.

\subsection{Energy conservation}
Energy conservation follows from the timelike projection of the conservation law,  $u_{\nu} \partial_{\mu} T^{\mu\nu}=0$, and yields
\begin{align}
\label{eq:energyconservation}
- d e_\Sigma - (e_\Sigma+p_\Sigma) \partial_\mu u^\mu =  - u_{\nu}\partial_{\mu}[  \partial^{\mu}\phi \cdot \partial^{\nu}\phi]+u_{\nu} \partial_{\mu}[
u^\mu u^\sigma u^\nu u^\rho \partial_{\rho}\phi \cdot \partial_{\sigma}\phi
] \, .
\end{align}
To simplify the notation, we introduce the shorthand 
\begin{equation}
\xi^\mu_a = \partial^{\mu}\phi_a , \quad \xi^\mu_a =-d\phi_a \, u^\mu + \partial_\perp^\mu \phi, 
\end{equation}
and then rhs of \ref{eq:energyconservation} can be rewritten as 
\begin{align}
u_{\nu} \partial_{\mu} (\xi^\mu\cdot \xi^\nu- u^{\mu}u^{\nu} (d\phi)^2  )
&= d\phi \cdot \partial_{\mu}\xi^\mu +\frac12 d \xi^2+
u_{\nu}  \xi^\mu\cdot( \partial_{\mu} \xi^\nu- \partial^{\nu}\xi_\mu)- u_{\nu}  \partial_{\mu}(u^{\mu}u^{\nu} (d\phi)^2   ). 
\end{align}
The curl vanishes due to the definition of $\xi$, and 
then using $d \xi^2 =d( d\phi)^2 + d  (\partial_\perp \phi^2)$
 this evaluates to
\begin{align}
   u_{\nu} \partial_{\mu} (\xi^\mu\cdot \xi^\nu- u^{\mu}u^{\nu} (d\phi)^2  )
= d\phi \cdot  \partial_{\mu} \partial_\perp^\mu \phi
+\frac12 d\,(\partial_\perp \phi)^2 \, .
\end{align}
Including the dissipative part of the energy-momentum tensor,  energy conservation yields finally  
  \begin{align}\label{energy-cons}
 d e_\Sigma + (e_\Sigma + p_\Sigma) \, \partial u =d\phi\, \cdot \partial_{\mu} 
 \partial^{\mu}_\perp \phi+ \frac12 d\,(\partial_\perp\phi)^2
+u_{\nu}\, \partial_{\mu}\Pi^{\mu\nu} \, .
\end{align}

\subsection{Charge Conservation }
The equation of (partial) current conservation reads  
\begin{align}
\partial_{\mu} J^{\mu}_{ab} 
 =   \phi_{a }H_{b }-\phi_{b }H_{a },
\end{align}
where the current is defined as 
\begin{equation}
J^{\mu}_{ab} =  
n_{ab} u^{\mu } + J^{\mu}_{\perp ab}+ q^{\mu}_{ab}.
\end{equation}
Here $n_{ab}$ is the charge,  $ J^{\mu}_{\perp ab}$ is the superfluid current in \eqref{current-super},  and $q^{\mu}_{ab}$ is the dissipative part of the current,   $q^{\mu}_{ab}u_{\mu}=0$. 
We then contract the eom with the antisymmetric tensor $\mu_{ab}$ and find
\begin{align}
   -\frac{1}{2} \mu_{ab}  \, ( dn_{ab } +n_{ab }\, \partial u)=
   \frac{1}{2} \mu_{ab} \, \partial_{\mu} q^\mu_{ab}
   + \frac{1}{2} \mu_{ab}\, \partial_{\mu} J^\mu _{\perp ab} + \mu_{ab} \,  \phi_{b }H_{a} \, .
\end{align}
Using the superfluid current in \eqref{current-super}, we find finally
\begin{align}\label{current-eq}
   -\frac{1}{2} \mu_{ab} \, ( dn_{ab } +n_{ab }\, \partial u)=
   \frac{1}{2} \mu_{ab} \,  \partial_{\mu} q^\mu_{ab} 
   +  \mu_{ab} \phi_b \left( \partial_{\mu} \partial_\perp^{\mu} \phi_a   - \frac{\partial V}{\partial \phi_a }  + H_a\right) \, , 
\end{align}
where we have inserted, 
$\phi_b\, \partial V/\partial \phi_a - \phi_a\,  \partial V/\partial\phi_b$, 
which vanishes due to the $O(4)$ symmetry of the potential.

\subsection{Synthesis}

After substitutions using \eqref{energy-cons} and \eqref{current-eq}, we find the final expression for the entropy production quoted in the text
\begin{align} 
   \partial_\mu(s_\Sigma u^{\mu} - \frac{\mu}{2T} \cdot q^{\mu}) 
   =&\frac{1}{T}\left(d\phi_a + \mu_{ab} \phi_b\right)  \, [\partial_{\mu} \partial^\mu_{\perp}\phi_a -\frac{\partial V}{\partial \phi_a} +H_a  ]
%
-\Pi^{\mu\nu}\partial_{\mu}\beta_\nu
- q^\mu \cdot\partial_{\mu}  \left( \frac{\mu}{2T} \right) \, .
\end{align}

\section{Computing the transport coefficients near the critical point}\label{app-exact}

In this appendix, we gather the details of the computation of the transport coefficients. First, we note that the dimensionless function introduced in \eqref{ffunc} can be integrated exactly
\begin{align}
   f_n(r,u) 
 &=\frac{16}{15\pi}\frac{ m^{7-2n}}{r^2-u^2 }
 \int_0^{\infty} dk\frac{k^{2n}}{(k^2 + m^2)^3} \left[ \frac{r^2}{ k^2 + r^2 m^2 }-\frac{u^2}{k^2 + u^2 m^2} \right],\\
 &=\frac{\sec (\pi  n) }{15\left(r^2-u^2\right)}
 \Big{[}\frac{4 n^2 \left(r^2-1\right)^2-8 r^{2 n+1}+8 n \left(r^2-1\right)-r^4+6
   r^2+3}{\left(r^2-1\right)^3} \nonumber\\
   & \qquad \qquad \qquad -\frac{4 n^2 \left(u^2-1\right)^2-8 u^{2 n+1}+8 n \left(u^2-1\right)-u^4+6
  u^2+3}{\left(u^2-1\right)^3}\Big{]}.
   \end{align}


Next, we take a closer look at the shear viscosity computation. We see from \eqref{shear-int} that the shear viscosity will have a contribution from the $\sigma$ and $\varphi$ propagators:
\begin{align}
\langle T^{xy}(x) T^{xy}(z) \rangle
 &= \langle \partial^x\delta\sigma(x)\partial^y\delta\sigma(x)  \partial^x\delta\sigma(z)\partial^y\delta\sigma(z) \rangle 
+ \sigma_0^4\langle \partial^x\varphi_a(x)\partial^y\varphi_a(x)  \partial^x\varphi_b(z)\partial^y\varphi_b(z)  \rangle, \nonumber\\
&\equiv I_{\sigma\sigma}^{xy}+I_{\varphi\varphi}^{xy} \, .
\end{align}
The contribution from the $\sigma\sigma$ propagator reads
\begin{align}
I_{\sigma\sigma}^{xy}
&= \frac{2 T^2}{(30\pi^2 \Gamma)} \int_0^{\Lambda}  \frac{k^6 dk  }{(k^2 + m^2_\sigma)^3}=
\frac{ T^2 \Lambda}{15\pi^2 \Gamma}  - \frac{  T^2 m_{\rm \sigma} }{16\pi \Gamma}.
\end{align}

Similarly, we need to evaluate the contribution to the shear viscosity due to the $\varphi\varphi$ propagator:
\begin{align}
I^{xy}_{\varphi\varphi} =2d_A \int \frac{d^3  k}{(2\pi)^3}\frac{d  \omega}{(2\pi)}\frac{\bsigma^4}{\omega_k^4} (k^x k^y G^{\varphi\varphi}_{\rm sym})^2
= 2 T^2 d_A \int \frac{d^3k }{(2\pi)^3}
\frac{(k^x k^y)^2}{ (k^2 + m^2)^2}  
\frac{g_2^2 +  (g_1g_2 + \omega_k^2)}{(g_1 + g_2) (g_1 g_2 + \omega_k^2) } \, .
\end{align}
We can evaluate the expression neatly by adding and subtracting the leading divergent piece
\begin{align}
I^{xy}_{\varphi\varphi} =  \frac{2T^2 d_A}{30\pi^2 \Gamma} \int_0^{\Lambda} \frac{k^6}{(k^2 + m^2)^3 }  
+   \frac{2 T^2}{30\pi^2} \int dk \frac{k^6}{(k^2 + m^2)^2} \left( 
\frac{g_2^2 +  (g_1g_2 + \omega_k^2)}{(g_1 + g_2) (g_1 g_2 + \omega_k^2) } - \frac{1}{g_1}  \right),\label{prev-int}
\end{align}
and by using \eqref{eq:u2def} and \eqref{eq:r2def}, we can 
evaluate the above expression to find
\begin{align}
I^{xy}_{\varphi\varphi} = \frac{2 T^2 d_A \Lambda}{30\pi^2 \Gamma}  - \frac{2 T^2 m d_A}{32 \pi \Gamma} \left(1 + 
u^2 (1 -r^2) f_3(r,u) \right).
\end{align}
Combining the ingredients, we find that the shear viscosity is given by \eqref{shearresult}.

\section{Comparison with pion kinetics}
\label{app:pionkinetics} 

Our purpose in this appendix is to explain the (black dashed) ``$\pi$-kinetics" curves
in \Fig{transport}.  As discussed in \Sect{transport-coeff},  when
writing down the hydrodynamic theory with the $\Sigma$ 
field we have integrated out modes with  $k\sim T$,
which  are then incorporated into the dissipative transport coefficients of the hydrodynamic theory such as $\eta_{\Sigma}$. 
Modes with $k \sim m_{\sigma}$ are explicitly propagated in the theory. 

At large negative $z$  (well in the broken phase),
the $\sigma$ is heavy  is compared to the pions, 
and can be consistently integrated out by  exploiting the mass hierarchy
\st
      m \ll m_\sigma \ll T \, . 
\stp
The resulting hydrodynamic effective theory  consists of energy, momentum, and light  pions, which are parameterized by the unitary matrix,  $U=e^{i2\varphi}$~\cite{Grossi:2020ezz}. 
Modes with $k\sim m_\sigma$ are  now incorporated into the new transport coefficients of this  theory such as $\eta_U$,  $\eta_U$ differs from $\eta_{\Sigma}$ due to the contribution of these  modes.  

At the longest distances with $k \ll m$,  the pion hydrodynamic theory reduces to ordinary hydrodynamics with the familiar transport coefficients $\eta$, $\zeta$ and $\sigma_I$. Matching the 
pion effective theory  to normal hydrodynamics determines the contribution
of soft pions to these normal  coefficients.
This computation gives~\cite{Grossi:2020ezz}
\begin{subequations}
   \label{eq:pioneft}
\begin{align}
   \eta  =& \eta_{U} -\frac{d_A T m}{120 \pi (\Gamma + D_0)}  \left[ \frac{ 2 r^3+4 r^2+6 r+3}{ (1+r)^2} \right],  \label{finaltransport2} \\         
\sigma_{I} =&    (\sigma_{I})_U  + 
\frac{T_A T}{24 \pi m (\Gamma + D_0)} \left[   \frac{1+ 2r}{(1 + r)^2 } \right]  ,\label{finaltransport3} \\
   \zeta =& \zeta_U- \frac{d_A T m}{8\pi (\Gamma + D_0)} \left(\frac{\beta c_{s0}^2 }{t}\right)^2 \left[ 
 \frac{8 r^3+16 r^2+16 r+7}{4 (1+r)^2 }
\right] . 
\end{align}
\end{subequations}
Here $r=\Gamma/(\Gamma + D_0)$, and $\eta_{U}$, $\zeta_{U}$, 
and $(\sigma_{I})_U$ are the dissipative parameters of the soft-pion effective theory\footnote{In \cite{Grossi:2020ezz}, the 
(renormalized) dissipative parameters  $\eta_{U}, \zeta_{U}, (\sigma_I)_U$  where called $\eta^{(0)}_{\rm phys}$, $\zeta^{(0)}_{\rm phys}$, and $(\sigma_I)^{(0)}_{\rm phys}$.  The transport coefficients 
$\Gamma$ and $D_0$ in this work were called $D_m$ and $D_A-D_m$ in
\cite{Grossi:2020ezz}. }.  

Expanding our results in \Eq{eq:transportresults} for
$\eta$ and $\sigma_I$ at large negative $z$ (where the parameter $u$ tends to infinity), we find that our expressions match with the pion EFT results in \eqref{eq:pioneft},  provided we identify
\begin{subequations}
   \label{eq:pioneft2}
\begin{align}
   (\sigma_I)_U&=  \sigma_\Sigma  +  \frac{T_A T}{12\pi \Gamma m_\sigma} \left( \frac{\Gamma + D}{D} \right) \left(\frac{\sqrt{\Gamma D m_\sigma^2} }{v} \right)_{-\infty} ,\\
   \eta_U &=  \eta_\Sigma  -\frac{T m_\sigma}{32\pi \Gamma}  - \frac{T d_A m_\sigma}{60\pi \Gamma}  \left(\frac{D}{\Gamma + D } \right)  \left(\frac{v}{\sqrt{\Gamma D m_\sigma^2}} \right)_{-\infty}. \label{etaU}
\end{align}
\end{subequations}
Here we have defined the constant
\st
\left(\frac{v}{\sqrt{\Gamma D m_\sigma^2 } } \right)_{-\infty} \equiv   \lim_{z\to-\infty}  \frac{v}{\sqrt{\Gamma D m_\sigma^2 } }   = \frac{u_c}{\sqrt{2}} \,  ,
\stp
where $u_c^2 \equiv v^2_c/\Gamma D m_c^2$ is a 
dimensionless combination of parameters  evaluated on the critical line (see \Sect{sec:linearresponse} for a physical explanation). Throughout the paper we have taken $u_c=1$ as in \Fig{fig:propagator}.  As discussed above, the  difference between $\eta_{U}$ and $\eta_{\Sigma}$ (and similarly for the conductivity) comes from integrating out modes with $k\sim m_{\sigma}$. Thus,  for instance, the second term in \eqref{etaU} stems from integrating out the $\sigma$ field, while the third term stems from integrating out hard pions with $k\sim m_\sigma$. The ``$\pi$-kinetics" curves in \Fig{transport}
are the asymptotics given in \eqref{eq:pioneft} with parameters identified in \eqref{eq:pioneft2}.

\end{appendix}

%


\end{document}